\documentclass[onecolumn,draftcls,12pt, twoside]{IEEEtran}

\usepackage{tabu}

\usepackage{amsmath,amsthm,graphicx,cite}
\usepackage{srcltx}
\usepackage{epsfig,amsfonts}
\usepackage{graphicx,cite,amssymb,amsmath}

\usepackage{amsthm}
\usepackage[nolist]{acronym}
\usepackage{psfrag}
\usepackage{perso}

\usepackage{pgfplots}
 \pgfplotsset{compat=newest}
    \pgfplotsset{plot coordinates/math parser=false}
    \pgfplotsset{
    label style={anchor=near ticklabel},
    xlabel style={yshift=0.0em},
    ylabel style={yshift=-0.3em},
    tick label style={font=\footnotesize },
    label style={font=\footnotesize},
    legend style={font=\footnotesize},
    title style={font=\fontsize{7}}}

\usepackage{relsize}
\usepackage{mathtools}
\mathtoolsset{showonlyrefs}

\usepackage{placeins}

\usepackage{multirow}

\usepackage{subfig}

\usepgflibrary{arrows}
\usetikzlibrary{patterns}
\usepgflibrary{decorations.pathmorphing}

\newcommand{\rippleset}{\mathscr{R}}
\newcommand{\cloudset}{\mathscr{C}}
\newcommand{\cloudsetc}[1]{\cloudset_{#1}}

\newcommand{\Ripple}{\mathtt{R}}
\newcommand{\Cloud}{\mathtt{C}}
\newcommand{\Cloudc}[1]{\Cloud_{#1}}

\renewcommand{\c}{\mathtt{c}}
\ifthenelse{\isundefined{\cc}}{\newcommand{\cc}[1]{\c_{#1}}}{\renewcommand{\cc}[1]{\c_{#1}}}

\renewcommand{\r}{\mathtt{r}}

\newcommand{\Ripp}{\mathsf{R}}

\newcommand{\ripple}[1]{ \msr{R}_{#1}}

\newcommand{\cloudc}[2]{\msr{C}_{#1,#2}}

\newcommand{\ru}{\r_u}
\newcommand{\Ru}{\Ripple_u}

\newcommand{\Cuc}[1]{\Cloud_{#1,u}}

\newcommand{\cuc}[1]{\c_{#1,u}}
\newcommand{\cucs}[2]{\c_{#1,#2}}

\renewcommand{\S}[1]{{\mathtt{S}_{#1}}}
\newcommand{\s}[1]{\pmb{\mathtt{s}}_{#1}}
\newcommand{\w}{\pmb{\mathtt{w}}}
\ifthenelse{\isundefined{\C}}{
    \newcommand{\C}[1]{\mathtt{C}_{#1}}
    }{
    \renewcommand{\C}[1]{\mathtt{C}_{#1}}
    }

\ifthenelse{\isundefined{\N}}{
    \newcommand{\N}[1]{\mathsf{N}_{#1}}
    }{
    \renewcommand{\N}[1]{\mathsf{N}_{#1}}
    }

\newcommand{\Erv}{\mathtt{A}}
\newcommand{\erv}{\mathtt{a}}
\renewcommand{\b}{\mathtt{b}}
\newcommand{\B}{\mathtt{B}}
\newcommand{\n}{\mathtt{n}}

\renewcommand{\nu}{\n_u}

\newcommand{\bh}[1]{\b_{#1}}
\newcommand{\buc}[1]{\b_{#1,u}}

\newcommand{\Buc}[1]{\B_{#1,u}}

\newtheorem{mydef}{Definition}

\newtheorem{theorem}{Theorem}

\newcommand{\myop}[1]{%
  \mathchoice{\raisebox{8pt}{$\displaystyle #1$}}
             {\raisebox{8pt}{$#1$}}
             {\raisebox{4pt}{$\scriptstyle #1$}}
             {\raisebox{1.6pt}{$\scriptscriptstyle #1$}}}

\newcommand{\paccess}{p}
\newcommand{\paccessc}[1]{\paccess_{#1}}

\newcommand{\betac}[1]{\beta_{#1}}

\newcommand{\p}{q}

\newcommand{\puc}[1]{\p_{#1,u}}

\newcommand{\slot}{y}

\newcommand{\per}{{\mathsf{P}}}
\newcommand{\throughput}{\mathsf{T}}
\newcommand{\nuser}{n}
\newcommand{\nslot}{m}
\newcommand{\nslotc}[1]{\nslot_{#1}}

\newcommand{\cs}{k}

\renewcommand{\deg}{\text{deg}}

\renewcommand{\P}{P}
\newcommand{\Pu}{P_u}
\newcommand{\x} {x}
\newcommand{\xc}[1]{{\x_{#1}}}

\newcommand{\y} {y}

\newcommand{\pccru} {p_{\cc{1},\cc{2}, \r,u} }
\newcommand{\pccrn} {p_{\cc{1},\cc{2}, \r,\nuser} }
\newcommand{\pccrrn} {p_{\cc{1},\cc{2}, \r_1 + \r_2,\nuser} }
\newcommand{\pccruprime} {p_{\cc{1}',\cc{2}', \r',u-1} }
\newcommand{\pcccru} {p_{\cc{1},\cc{2}, \hdots, \cc{\cs} \r,u} }
\newcommand{\pcccrn} {p_{\cc{1},\cc{2}, \hdots, \cc{\cs} \r,\nuser} }

\newcommand{\cho}{ \mathring{c}_{h} }
\newcommand{\ro}{ \mathring{r} }

\newcommand{\tp}{t}
\newcommand{\nslotT}{\nslot_T}
\newcommand{\nusernew}{u}
\newcommand{\nslotnew}{\nslot'}
\newcommand{\he}{k}

\newcommand{\St}{\mathsf{I}}
\newcommand{\st}{\mathtt{i}}
\newcommand{\J}{\mathsf{J}}
\renewcommand{\j}{\mathtt{j}} 

\newcommand{\validstates}{\mathcal{S}}

\usepackage{multirow}

\begin{document}
\begin{acronym}
\acro{AP}{access point}
\acro{RA}{random access}
\acro{LT}{Luby Transform}
\acro{BP}{belief propagation}
\acro{i.i.d.}{independent and identically distributed}
\acro{IoT}{Internet of things}
\acro{MTC}{machine type communication}
\acro{PER}{packet error rate}
\acro{SIC}{successive interference cancellation}
\acro{URLLC}{ultra-reliable low latency communication}
\acro{PMF}{probability mass function}
\end{acronym}


\title{Reliability-Latency Performance of Frameless ALOHA with and without Feedback}

\author{
    \IEEEauthorblockN{Francisco L\'azaro\IEEEauthorrefmark{1}, \v Cedomir Stefanovi\'c\IEEEauthorrefmark{2},  Petar Popovski\IEEEauthorrefmark{2}\\
     \IEEEauthorblockA{\IEEEauthorrefmark{1}Institute of Communications and Navigation of DLR (German Aerospace Center),
    \\Wessling, Germany. Email: Francisco.LazaroBlasco@dlr.de}\\
     \IEEEauthorblockA{\IEEEauthorrefmark{2}Department of Electronic Systems, Aalborg University
    \\Aalborg, Denmark. Email: \{cs,petarp\}@es.aau.dk}\\
\thanks{This work has been presented in part at ITG SCC, Hamburg, Germany, February 2017 \cite{lazaro:SCC2017}, and IEEE Globecom, Singapore, December 2017 \cite{stefanovic:globecom2017}.}
\thanks{This work of Petar Popovski has been in part supported by the Danish Council for Independent Research (Grant Nr. 8022-00284B SEMIOTIC).}
\thanks{This work has been accepted for publication in IEEE Transactions on Communications.}
\thanks{\copyright 2020 IEEE. Personal use of this material is permitted. Permission from IEEE must be obtained for all other uses, in any current or future media, including
	reprinting /republishing this material for advertising or promotional purposes, creating new
	collective works, for resale or redistribution to servers or lists, or reuse of any copyrighted
	component of this work in other works}
}
}

\maketitle

\thispagestyle{empty} \pagestyle{empty}

\vspace{-1.75cm}

\begin{abstract}
This paper presents a finite length analysis of multi-slot type frameless ALOHA based on a dynamic programming approach.
The analysis is exact, but its evaluation is only feasible for moderate number of users due to the computational complexity.
The analysis is then extended to derive continuous approximations of its key parameters, which, apart from providing an insight into the decoding process, make it possible to estimate the packet error rate with very low computational complexity.
Finally, a feedback scheme is presented in which the slot access scheme is dynamically adapted according to the approximate analysis in order to minimize the packet error rate.
The results indicate that the introduction of feedback can substantially improve the performance of frameless ALOHA.
\end{abstract}

\vspace{-0.5cm}

\section{Introduction}\label{sec:Intro}

Random access protocols find applications in scenarios where there is a number of users sharing a common transmission medium and there exists uncertainty regarding which users are active.
They can be used both in the initial phase of grant-based access, where the active users contend with metadata in order to reserve the uplink resources for the subsequent data transmissions, or in grant-free access, where the active users contend directly with packets containing data.
The former approach forms the basis of mobile cellular access, e.g.,~\cite{TS36.321}.
However, the latter approach has been gaining research momentum recently, e.g.,~\cite{R1-1808304}, due its lower signaling overhead which makes it suitable for systems such as \ac{IoT}, where the amount of exchanged data is small but the number of contending users may be large~\cite{Laya2014}.

The first and still widely used random access protocols are ALOHA and slotted ALOHA \cite{R1975}.
Assuming a collision channel model, these two protocols offer low peak throughput ($1/2e$ and $1/e$, respectively) and high \ac{PER} even for low channel load.
However, it has been shown how the introduction of \ac{SIC} at the receiver can lead to higher performance by leveraging on coding-theoretic tools \cite{CGH2007,L2011}.
In practice, this implies storing and processing the receiver waveform and leads to higher receiver complexity.
The results presented in \cite{L2011} inspired a strand of works that applied various concepts from codes-on-graphs to design SIC-enabled slotted ALOHA schemes~\cite{PLC2011,LPLC2012,SPV2012,PSLP2014,JBVC2015,SGB2017}, which are usually referred to by using the umbrella term of \emph{coded slotted ALOHA}.

Frameless ALOHA~\cite{SPV2012,SP2013} is a version of SIC-enabled slotted ALOHA that exploits ideas originating from the rateless-coding framework \cite{luby02:LT}.
In its original version, frameless ALOHA is characterized by a contention period that consists of a number of slots that is not defined a priori, and by a slot access probability for the users to independently transmit their packets in a slot.
An asymptotic optimization of the slot access probability that maximizes the expected throughput was performed in \cite{SPV2012}, while a similar optimization in a cooperative, multi-base station scenario was recently considered in \cite{OIA2018}.
In \cite{Ogata:2019}, an evolution of frameless ALOHA was proposed which makes use of feedback, ZigZag decoding, and a dynamical variation of the slot access probability  to improve performance.
A joint assessment of the optimal slot access probability and the contention termination criteria in non-asymptotic, i.e., finite-length scenarios were assessed by means of simulations in \cite{SP2013}.

The finite-length performance of frame-based coded slotted ALOHA schemes has been studied in \cite{ivanov:floor,SGB2017,graell:2018,fereydouniannon,MGS2018}. Here we remark that, due to the fact that in such schemes every user selects a random set of slots in a frame in which to transmit, interdependencies among slots are introduced; thus, the finite-length analytical model is in essence intractable, and one has to resort to approximations. The error floor of coded slotted ALOHA over the collision channel in the finite length regime was studied in \cite{ivanov:floor}. A frame asynchronous coded slotted ALOHA scheme was analyzed in \cite{SGB2017,graell:2018}. The scaling laws of coded slotted ALOHA were derived in \cite{fereydouniannon}. In \cite{MGS2018}, the performance of several advanced random access protocols was compared over the collision channel and over the additive white Gaussian noise channel.

The focus of this paper on finite-length analysis of frameless ALOHA and of its reliability-latency performance with and without feedback. In particular, the considered feedback takes the form of an update of the slot access probability of frameless ALOHA, which is the default degree of freedom that can be used to optimize performance.
	This feedback induces multiple slot types in the contention, which substantially expands the design space, but also poses some challenges in terms of modeling and analytical performance optimization.
	The analysis presented in the paper characterizes statistically the reliability of muli-slot type frameless ALOHA for a predefined latency target, which in our context translates to a fixed number of slots in the contention period. The motivation of this approach stems from the fact that mission-critical services typically feature a given latency budget, which is also reflected in the proverbial reliability requirement of $10^{-5}$ under the latency of 1~ms in the context of 5G ultra-reliable low-latency service category~~\cite{TR38.913}.\footnote{We also remark that extending indefinitely the contention period in frameless ALOHA will always be beneficial in terms of reliability; this is a general feature of the rateless-coding framework. However, a practical design target always involves limited latency, which in this case maps to limited number of
		overhead slots.}
	The feedback scheme proposed in this paper exploits the flexibility that frameless ALOHA offers with respect to the frame-based schemes. Specifically, the considered scheme assesses the reliability at predefined intermediate checkpoint within a contention period and uses feedback to drive the contention process towards maximizing the number of decoded users, and thereby the reliability, given the latency target.
	In particular, the feedback sent by the access point to the users consists of an update on the slot access probability that should be subsequently used. This feedback induces a multi-slot type contention, since a new slot type is created whenever the slot access probability is changed.
	The proposed scheme adapts to the reliability-latency performance observed at multiple points, which allows for a finer control of the contention process.
	This paradigm change is a clear indicator of an enlarged design space compared to the frame-based designs that focus on just a single time instant, placed at the end of the latency budget. 
	Finally, we show that the proposed feedback scheme outperforms frame-based schemes. 

This paper builds on preliminary results from \cite{lazaro:SCC2017} and \cite{stefanovic:globecom2017}. In \cite{lazaro:SCC2017} an exact finite-length analysis of frameless ALOHA  was presented for the case in which all slots are statistically identical.
This analysis was then extended in  \cite{stefanovic:globecom2017} to multiple slot types, sketching how the analysis could be changed to accommodate this extension without actually providing a full proof.
The contributions in this paper are the following.
We present in detail an exact, self-contained finite-length analysis of multi-slot type frameless ALOHA. Due to computational complexity, the analysis is only feasible for moderate number of users. This has been the motivation to extend the analysis towards deriving continuous approximations of the expected ripple size (the number of slots containing only one transmission) and its standard deviation.
Based on these approximations, we propose a method to estimate the \ac{PER} with very low complexity.
Finally, we exploit this estimation of the packet error rate to propose a feedback based scheme in which the slot access probability of frameless ALOHA is adapted dynamically.\footnote{Note that, although the paper ultimately proposes the approximate method in order to deal with the computational complexity, the nature of the approximation is substantially different from the approximations employed in \cite{ivanov:floor,SGB2017,graell:2018,fereydouniannon,MGS2018}. Specifically, in our case, the exact model and analysis of the problem are at hand, as shown in the paper, and one can, in principle, exactly assess the quality of the approximation. In case of the methods employed in \cite{ivanov:floor,SGB2017,graell:2018,fereydouniannon,MGS2018}, the analysis is from the very beginning approximate due to the interdependencies among slot selections by the users.}
We investigate the performance of the dynamic version of the scheme and show that it achieves a rather favorable performance, which progressively improves with the number of feedback opportunities at the expense of an increased computational burden.

The remainder of the paper is organized as follows.
Section~\ref{sec:II} provides a brief overview of frameless ALOHA and describes the system model.
Section~\ref{sec:analysis} presents the finite-length analysis, which can be used to obtain the exact probability mass function of the number of unresolved users for a given duration of the contention period.
In Section~\ref{sec:diff_eqs}, a low complexity approximate analysis of the frameless ALOHA decoder is presented, which is then used to estimate  the packet error rate.
Section~\ref{sec:dynamic} shows how it is possible to largely improve the performance of frameless ALOHA by introducing feedback and applying the analysis derived in this paper.
Finally, Section~\ref{sec:conclusions} concludes the paper.


\section{Background and System Model}
\label{sec:II}

\subsection{Background: Frameless ALOHA}
\label{sec:background}

\begin{figure}[t]
        \centering
        \includegraphics[width=0.85\columnwidth]{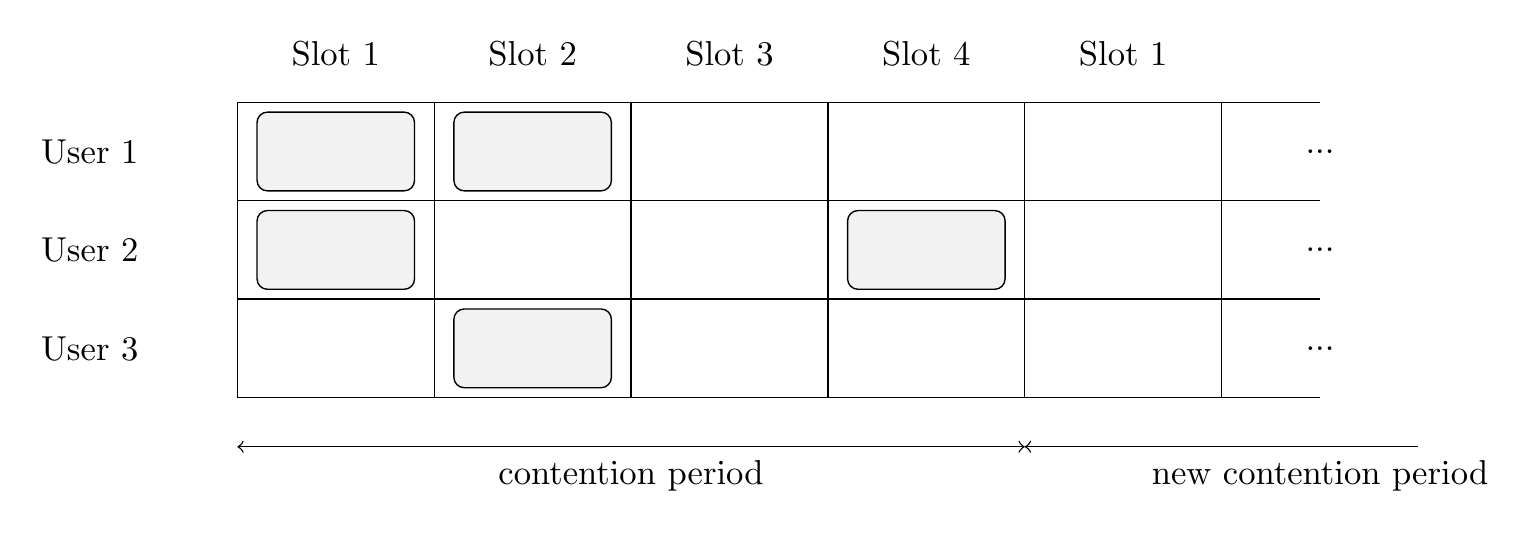}
        \caption{An example of contention in Frameless ALOHA. All three users randomly and independently decide on a slot basis whether to transmit or not. Slot 1 and slot 2 are collision slots; the colliding transmissions can not be decoded and the \ac{AP} stores the slots (i.e., the signals observed in them) for later use. Slot 4 is a singleton slot and the \ac{AP} decodes a replica of the packet of user 2 from it. The \ac{AP} also learns that a replica of packet of user 2 occurred in slot 1, and removes (cancels) it from the stored signal. Slot 1 now becomes singleton and a replica of the packet of user 1 becomes decoded. In the same manner, the successive process of replica removal and decoding of a new packet replica occurs in slot 2. As all three users have become resolved, the \ac{AP} terminates the contention period after slot 4, and starts a new one.}
        \label{fig:frameless_example}
\end{figure}

Frameless ALOHA \cite{SPV2012} can be regarded as a variant of slotted ALOHA with \ac{SIC} that is inspired by rateless codes \cite{luby02:LT}.
The time in frameless ALOHA is divided into equal-length slots and slots are organized into contention periods, whose length is a-priori not known.
In order to transmit a packet, users must first wait until a new contention period starts.
Next, in each slot of the contention period, every contending user transmits a replica of its packet with a predefined slot-access probability. This happens independently from the transmission in other slots and independently of the actions of any other contending user.
Furthermore, the assumption is made that each packet replica contains information about the slots in which the other replicas of the same packet are placed.
This could be accomplished by, e.g., enriching the packet header with the seed of a random number generator and the transmit decisions are determined by the output of this generator. Hence, when a packet replica is decoded, it provides information about the timing of all other replicas.
The  \acf{AP}  is required to store the waveform of the whole contention period and processes slots sequentially, leveraging \ac{SIC} to remove the interference caused by replicas of decoded packets. Since every packet contains a pointer to the location of all its replicas,  when a packet is successfully decoded, the receiver can determine the location of all its replicas and remove them from the received waveform. This reduces the interference in those slots containing replicas of the decoded packets and it may enable the receiver to decode more packets within those slots. 
 This process is repeated until no more packets can be decoded.
 
This \ac{SIC} process is not exclusive to frameless ALOHA, but common to all SIC-enabled slotted ALOHA schemes. The peculiarity of frameless ALOHA is the fact that the contention period (also known as frame) duration need not be a priori defined. The \ac{AP} can trigger the \ac{SIC} process after  the reception of every slot, and decide whether to end the contention period and start a new one, or alternatively to let the contention period continue. 
This decision is made according to a predefined criterion, e.g., whether the target throughput has been reached and/or a predefined fraction of users have been resolved \cite{SP2013}.
The start or termination of a contention period can be signaled to users by means of a beacon signal transmitted by the access point \cite{SP2013}.
An example of contention period in frameless ALOHA is depicted in Fig.~\ref{fig:frameless_example}.

\subsection{System Model}
\label{sec:sysmodel}

We denote by $\nuser$ the number of users contending for access to a single access point.
The duration of the contention period in slots is denoted by $\nslot$; note that  $\nslot$ is not a-priori fixed.
Furthermore, we shall assume that $\cs$ different slot types exists.
In particular, we assume that, out of the total $\nslot$ slots, exactly $\nslotc{1},  \nslotc{2}, ... \nslotc{\cs}$ are of type $1,2, ...,\cs$. 
Slots of type $h$, are characterized by a slot access probability $\paccessc{h}$, given by $\paccessc{h} = \frac{\betac{h}}{\nuser}$, which is equal for all users. 
Hence, in a slot of type $h$, every contending user will be active (send a packet) with probability $\paccessc{h}$, independently from its transmission in others slots, and from the actions of the other users.
It is easy to verify that $\betac{h}$ is the mean number of users that transmitted in a slot of type $h$, and, thus, equal to the expected number of transmissions contained in the slot.

\begin{figure}[t]
        \centering
        \includegraphics[width=0.63\columnwidth]{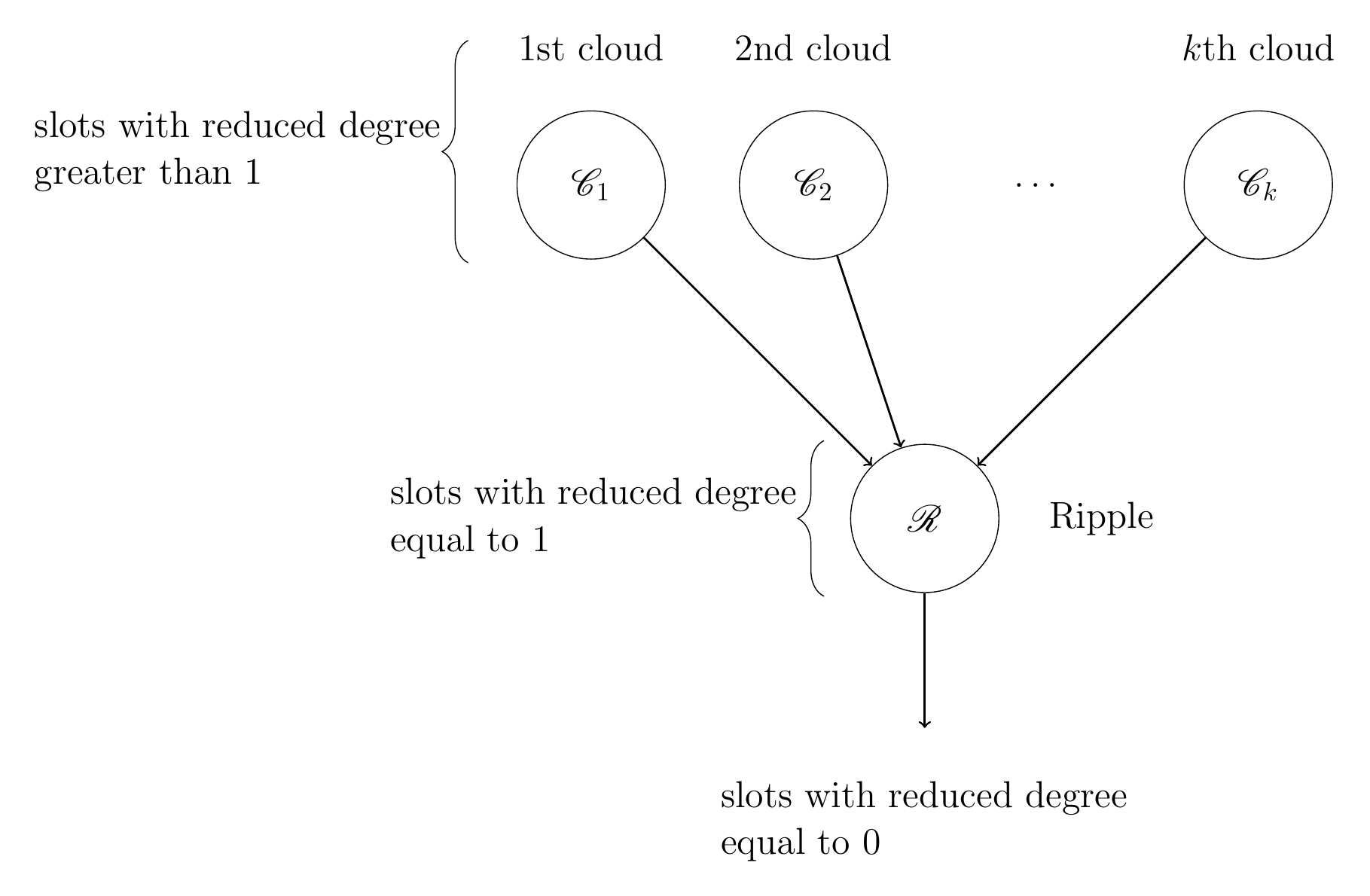}
        \caption{The clouds and the ripple.} 
        \label{fig:ripple_cloud}
\end{figure}

A collision channel model will be assumed. Hence, singleton slots, i.e., slots containing a single transmission, are decodable with probability $1$, and collision slots, i.e., slots containing two or more transmissions, are not decodable with probability $1$.
Perfect interference cancellation will be assumed, i.e., the removal of replicas from the slots leaves no residual transmission power.\footnote{This assumption is reasonable for practical interference cancellation methods and moderate to high signal-to-noise ratios \cite{L2011}.}

In order to model the successive interference cancellation process at the receiver, we introduce the following definitions:

\vspace{-4mm}
\begin{mydef}[Initial slot degree] The initial slot degree is the number of transmissions originally occurring in the slot.
\end{mydef}

\vspace{-6mm}
\begin{mydef}[Reduced slot degree] The reduced slot degree is the current number of unresolved transmissions in the slot, over the iterations of the reception algorithm.
\end{mydef}

\vspace{-6mm}
\begin{mydef}[Ripple] The ripple is the set of slots of reduced degree 1, and it is denoted by $\rippleset$.
\end{mydef}

\vspace{-4mm}
\noindent The cardinality of the ripple, $|\rippleset|$ is denoted by $\r$ and its associated random variable as $\Ripp$. 

\vspace{-4mm}
\begin{mydef}[$h$-th cloud] The $h$-th cloud, $\cloudsetc{h}$, is the set of slots of type $h$ with reduced degree $d > 1$.
\end{mydef}

\vspace{-4mm}
\noindent The cardinality of the $h$-th cloud, $| \cloudsetc{h}| $,  is denoted by $\cc{h} $ and the corresponding random variable as $\Cloudc{h}$.

Upon reception, the reduced degree of a slot is equal to its initial degree, since all users that are active in the slot are unresolved. During the decoding process, the reduced degree of a slot is decreased by $1$ whenever one of the unresolved users which are active in the slot is decoded and its interference is cancelled. 
Whenever the reduced degree of a slot in the $h$-th cloud becomes $1$, the slot leaves the $h$-th cloud and enters the ripple. Note, however, that when the reduced degree of a slot decreases from, say, $3$ to $2$, then the slot remains in the $h$-th cloud. Similarly, when the only unresolved user in a slot belonging to the ripple is resolved, the reduced degree of the slot becomes $0$ and the slot leaves the ripple. The slots of reduced degree $0$ are not of further use in the analysis, and are thus not considered explicitly. The transition of slots between the clouds and ripple is depicted in Fig.~\ref{fig:ripple_cloud}.

In order to keep track of the temporal variation of the ripple and cloud sizes as the \ac{SIC} process progresses, we introduce a subscript $u$, which corresponds to the number of unresolved users. Thus, initially we have $u=\nuser$, and every time a user is resolved, $u$ is decreased by 1 until we have decoded all users, when $u=0$. Thus, the ripple when $u$ users are unresolved is denoted by $\ripple{u}$, its cardinality by $\r_{u}$ and the random variable associated to it by $\Ripp_{u}$.  Similarly, when $u$ users are unresolved,  $\cloudc{h}{u}$, $\cuc{h}$ and $\Cloudc{h,u}$ denote the $h$-th cloud, its cardinality and the associated random variable, respectively.

Let us consider the example in Fig.~\ref{fig:frameless_example}, and assume that there are two slot types. Further, let us assume that slot 1 and 3 are of type 1, whereas slot 2 and 4 are of type 2. Initially slots 1 and 2 have degree $2$, hence they belong to the $1$-st cloud  $\cloudsetc{1}$ and  $2$-nd cloud $\cloudsetc{2}$, respectively. Similarly, slot 4 is initially in the ripple, $\rippleset$, since it has a degree of one. In the first decoding step, the only active user in slot 4, user 2, gets resolved and the replica transmitted by user 2 in slot 1 is cancelled. As a consequence, slot 4 leaves the ripple, while slot 1 leaves the $1$-st cloud and enters the ripple. Hence, the $1$-st cloud becomes empty.
In the next step, user 1 is resolved, slot 1 leaves the ripple, and slot 2 leaves the $2$-nd cloud and enters the ripple. Thus, the $2$-nd cloud becomes empty.
Finally, user 3 is resolved, and slot 2 leaves the ripple. In this case, all 3 users could be resolved.

Finally, we denote the slot degree distribution of  slot type $h$ by $\mathbf{\Omega_{h} }=\left\{ \Omega_{h,1}, \Omega_{h,2},...,\Omega_{h,\nuser} \right\}$, $h=1,2,...,\cs$, where $\Omega_{h,j}$ is the probability that a slot of type $h$ has initial degree $j$.
It is straightforward to verify that $\Omega_{h,j}$, $j=1,2,...,\nuser$, is given by
\begin{equation}\label{eq:omega_h_j}
\Omega_{h,j} = { \nuser \choose j } \left(\paccessc{h} \right)^j \left( 1 - \paccessc{h} \right)^{\nuser - j} 
 = { \nuser \choose j }\left( \frac{\betac{h}}{\nuser} \right ) ^j \left( 1 - \frac{\betac{h}}{\nuser} \right)^{\nuser - j}.
\end{equation}
Hence, the probability that a slot of type $h$ initially contains no transmission at all is $\Omega_{h,0}$, the probability that it initially belongs to the ripple is $\Omega_{h,1}$, and the probability that it initially belongs to the $h$-th cloud is $( 1 -  \Omega_{h,0} - \Omega_{h,1} ) $.


\section{Finite-Length Analysis}

For the sake of convenience and without loss of accuracy, we shall assume that the receiver works iteratively.
If the ripple is empty, the receiver simply stops. Otherwise, it carries out the following steps:
\begin{itemize}
\item Selects at random one of the slots in the ripple;
\item Resolves the user that was active in that slot (i.e., decodes its packet);
\item Cancels the interference contributed by the resolved user from all other slots in which its packet replicas were transmitted. This may cause some slots to leave the cloud and enter the ripple. Furthermore, some slots from the ripple  may become degree zero and leave the ripple. These last slots correspond to the slots in the ripple in which the resolved user was active.
\end{itemize}
Thus, in each iteration, the reception algorithm either fails, or exactly one user gets resolved.
These assumptions are made to ease the analysis and have no impact on the performance.

\label{sec:analysis}

Following the approach in \cite{Karp2004,lazaro:inact2017,lazaro:SCC2017}, the iterative reception of frameless ALOHA  with $\cs$ different slot types is represented as a finite state machine with state
\[
\S{u} :=( \Cuc{1}, \Cuc{2}, \cdots, \Cuc{\cs}, \Ru )
\]
i.e., the state comprises the cardinalities from the first to  $\cs$-th cloud and the ripple at the reception step in which $u$ users are unresolved.
Each iteration of the reception algorithm corresponds to a state transition.
The following proposition establishes a recursion that can be used to determine the state distribution.
\begin{theorem}\label{theorem:class}
Given that the decoder is in state ${\S{u}=(\cuc{1}, \cuc{2}, \cdots, \cuc{\cs}, \ru )}$, where $u$ users are unresolved, and $\ru>0$ (i.e., the ripple is not empty), the probability of the receiver transitioning to state ${\Pr \{\S{u-1}= \s{u-1}\}}$, where $u-1$ users are unresolved, is given by
\begin{align}
\Pr & \{ \S{u-1}= (  \s{u} + \w )| \S{u} =  \s{u} \} =
\binom{\ru-1}{\erv_u-1} \left(\frac{1}{u}\right)^{\erv_u-1} \times \\
& \left( 1- \frac{1}{u} \right)^{\ru-\erv_u} \mathlarger{\prod}_{h=1}^{h=\cs} { \binom{\cuc{h}}{\buc{h}} {\puc{h}}^{\buc{h}} (1-\puc{h})^{\cuc{h}-\buc{h}}  }
\label{eq:prob_transition}
\end{align}
with
\[\s{u} =(\cuc{1}, \cuc{2}, \cdots, \cuc{\cs}, \ru ) \]
\[{\w =(-\buc{1}, -\buc{2}, \cdots , -\buc{\cs}, \sum_{h=1}^{\cs} \buc{h} - \erv_u ) }\]
and
\vspace{-2ex}
\begin{align}
\puc{h} =  \frac{ \mathlarger {\sum}\limits_{d=k+1}^{\nuser}   \Omega_{h,d} \, \frac{d}{\nuser}  \binom{d-1}{\cs}  \frac{\binom{u-1}{\cs}}{\binom{\nuser-1}{\cs}}  \frac{\binom{\nuser-u}{d-\cs-1}}{\binom{\nuser-\cs-1}{d-\cs-1}} }
{  1 -  \mathlarger{\sum}\limits_{h=0}^{\cs} \,\,
\mathlarger{\sum}\limits_{d=h}^{\nuser}    \Omega_{h,d} \, \frac{\binom{u}{h} \binom{\nuser-u}{d-h}}{\binom{\nuser}{d}} }
\label{eq:pu_theorem_class}
\end{align}
for $0 \leq \buc{h} \leq \cuc{h}$,  $1 \leq \erv_u \leq \ru$.
\end{theorem}

\begin{IEEEproof}
 See Appendix~\ref{app:proof_theorem_class}.
\end{IEEEproof}

Recall that out of the $\nslot$ slots in the contention period, exactly $\nslotc{1},  \nslotc{2}, ... \nslotc{\cs}$ belong to slot type $1,2, ...,\cs$.
We focus on slots of type $h$, of which there are $\nslotc{h}$.
The initial state distribution corresponds to a multinomial with $\nslotc{h}$ experiments (slots) and three possible outcomes for each experiment: the slot being in the cloud, the ripple or having degree 0, with respective probabilities $(1-\Omega_{h,1} - \Omega_{h,0} )$, $\Omega_{h,1}$ and $\Omega_{h,0}$.
Denoting by  $\Ripple_{h, \nuser}$ the random variable associated to the number of slots of type $h$ of reduced degree 1 when all $\nuser$ users are still undecoded, we have
\begin{align}\label{eq:init_cond_finite}
 \Pr\{(\Cloud_{h, \nuser} = \c_{h,\nuser}, \Ripple_{h,\nuser} = \r_{h, \nuser} \} &=
 \frac{\nslotc{h}!}{\c_{h,\nuser}! \, \r_{h,\nuser}!\, (\nslotc{h}-\c_{h,\nuser}-\r_{h,\nuser})!}  \\
 &\times  \left( 1-\Omega_{h,1}- \Omega_{h,0}\right)^{\c_{h,\nuser}}\, {\Omega_{h,1}}^{\r_{\nuser}} \, {\Omega_{h,0}}^{\nslotc{h}-\c_{h,\nuser}-\r_{h,\nuser}}
\end{align}
for all non-negative $\c_{h, \nuser},\r_{h, \nuser}$ such that $\c_{h, \nuser}+\r_{h, \nuser} \leq \nslotc{h}$.

If we observe that, when all $\nuser$ users are still undecoded, the total number of degree one slots  $\Ripple_{\nuser}$ is given by
\[
\Ripple_{\nuser} = \sum_{h=1}^{\cs} \Ripple_{h, \nuser}
\]
we can obtain from \eqref{eq:init_cond_finite} the initial state distribution of the receiver.

By applying recursively Theorem~\ref{theorem:class}  and initializing as described the finite state machine, one obtains the state probabilities.

Let us denote by ${\per}_u$ the probability that exactly $u$ users remain unresolved after a contention period of $m$ slots. Obviously, the event that exactly $u$ users remain unresolved corresponds to the event that the user resolution ends at stage $u$.
 The probability of this event is simply the probability that the ripple is empty when $u$ users are still unresolved. Formally we have
\begin{align}
\label{eq:peru}
{\per}_u &= \Pr\{ \Ru =0\} = \sum_{\cuc{1}}  \sum_{\cuc{2}} \hdots \sum_{\cuc{\cs}} \Pr\{\S{u} =(\cuc{1},\cuc{2},\cdots, \cuc{\cs}, 0) \}
\end{align}
where the summations is taken over all possible values of $\cuc{h}$, $h=1, ..., \cs$.

Thus, by applying Theorem~\ref{theorem:class} and then using \eqref{eq:peru}, one obtains the \ac{PMF}  of the number of unresolved users, given number of users $\nuser$ and the duration of the contention period\footnote{Note that ${\per}_u$ implicitly depends on the initial state distribution that is obtained through \eqref{eq:init_cond_finite}, while \eqref{eq:init_cond_finite} depends on the number of slots of a given type $\nslot^{(h)}$, $h=1,2,...,\cs$, and thereby on the total number of slots $\nslot$.} $\nslot$.

The expected packet error rate $\per$ , i.e., the probability that a user is not resolved, can also be derived from Theorem~\ref{theorem:class}. In particular, we have
\begin{align} 
\per &= \sum_{u=1}^{\nuser} \frac{u}{\nuser}  {\per}_u 
=  \sum_{u=1}^{\nuser} \sum_{\cuc{1}}  \sum_{\cuc{2}} \hdots \sum_{\cuc{\cs}} \frac{u}{\nuser}\Pr\{\S{u} =(\cuc{1},\cuc{2},\cdots, \cuc{\c}, 0) \} \label{eq:per}.
\end{align}
Hence, the expected throughput is simply the ratio of the expected number of resolved users over the number of slots in the contention period. i.e.,
\begin{align}
\throughput &= \frac{\nuser ( 1-\per ) }{ \nslot }.
\end{align}

\begin{figure}[t]
        \centering
        \includegraphics[width=0.6\columnwidth]{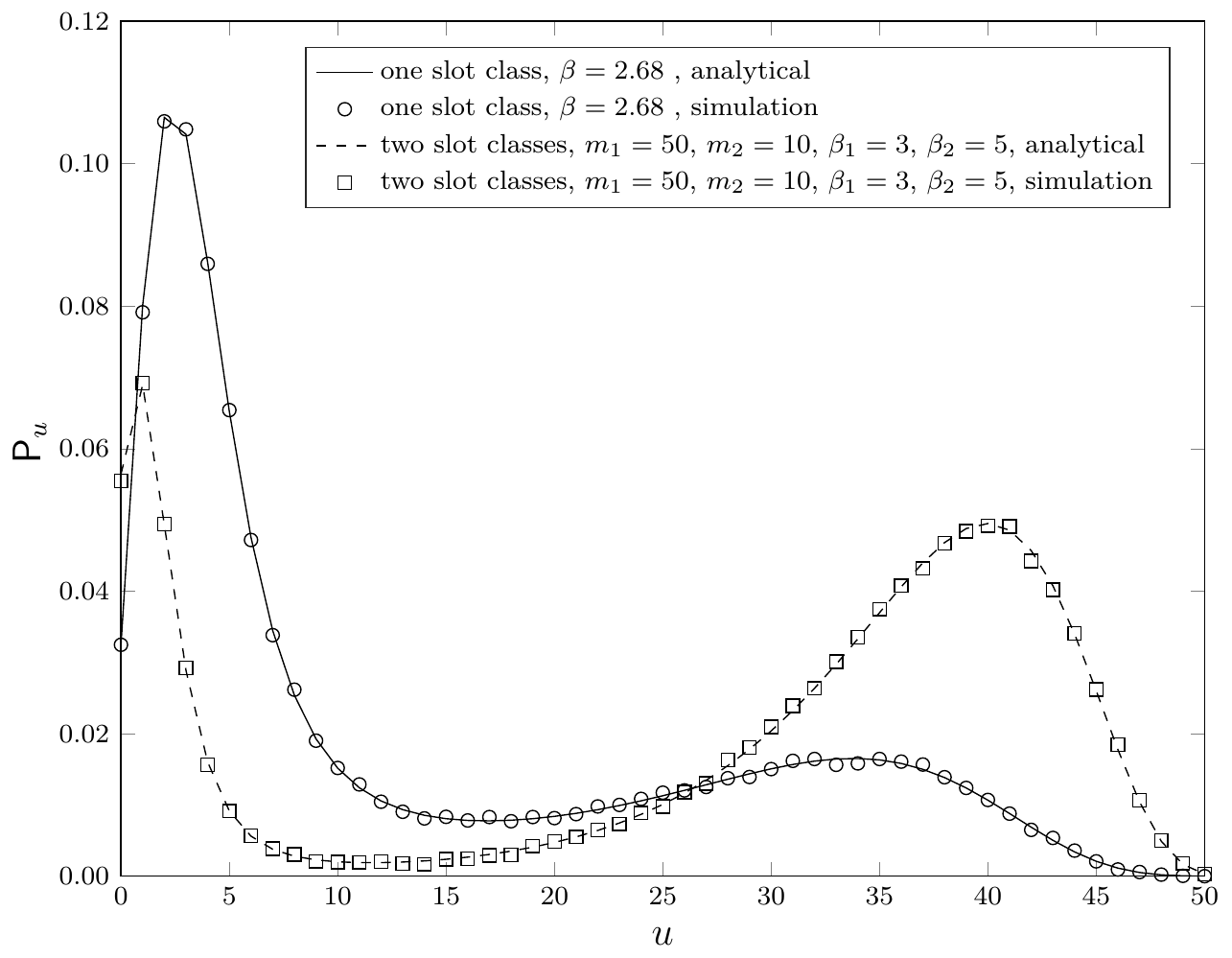}
        \caption{Examples of probability mass function of the number of undecoded users $u$ for $n=50$, $\nslot = 60$.}
        \label{fig:example_p_dist}
\end{figure}

As an example, in Fig.~\ref{fig:example_p_dist} we show the \ac{PMF} of the number of undecoded users $u$, i.e., $\per_u$ for $u=1,...,\nuser$, when $n=50$ and $\nslot = 60$, for frameless ALOHA with (i) a single slot type with mean initial slot degree $\beta=2.68$, and (ii) two slot types, with $\nslotc{1}=50$ slots of the first type, $\nslotc{2}=10$ slots of the second type, with mean initial degrees $\betac{1}=3$ and $\betac{2}=5$, respectively.\footnote{Recall that slot access probability of a type $h$ is $\paccessc{h} = \betac{h} / n$.}
The figure shows analytical results according to Theorem~\ref{theorem:class} and the outcome of Monte Carlo simulations.
It can be observed that the match is tight down to simulation error (100,000 contention periods were simulated).
In this particular example, we can observe how $\per_u$ has a bimodal distribution.
Thus, there are two points in the decoding process in which the ripple has a higher probability of becoming empty. For the example in Fig.~\ref{fig:example_p_dist}, the expected packet error rates  for the contention with one and two slot classes correspond to $0.264$ and $0.555$, respectively. 

\section{Approximation Using Differential Equations}\label{sec:diff_eqs}

The analysis presented in Section~\ref{sec:analysis}  can provide the exact probability mass function of the number of unresolved users, however, its evaluation is feasible only for a moderate number of users and a small number of slot classes. In some situations, such as when running a computer search to optimize a parameter, it is preferable to have an approximate analysis which can be efficiently evaluated. 

Having this in mind, 
in this section we derive continuous approximations of the first and second moment of the ripple and clouds. These approximations are easy to evaluate and can be used to obtain an insight in the decoding process.
Furthermore, we show how these approximations can be used to estimate the packet error rate.


\subsection{State generating Functions}\label{sec:state_gen_fun}

Our starting point to derive approximations of the distribution of the ripple and clouds is writing down the state generating function of the frameless ALOHA decoder, which is simply the probability generating function\footnote{The probability generating function is a representation of the probability mass function 
using power series.} of the random variable associated to  the state of the frameless ALOHA decoder.  

Following the works in \cite{Karp2004,shokrollahi2009theoryraptor,Maatouk:2012}, which analyze the iterative LT decoding process, let us define the state generating function of the frameless ALOHA decoder as 
\begin{align}
\Pu (\xc{1},\xc{2},\dots, \xc{\cs},\y) & := 
\sum_{ \validstates } \pcccru \,  {\xc{1}}^{\cc{1}} \, {\xc{2}}^{\cc{2}} \dots {\xc{\cs}}^{\cc{\cs}}  \,\y^{r-1}.
\end{align}
where $\validstates$ represents the set of all valid decoder states, 
\begin{align} \label{eq:valid_states}
    \validstates = \Big\{ (\cc{1}, \cc{2}, \hdots, \cc{\cs}, \r) | \cc{1} \geq 0,  \cc{2} \geq 0, \hdots \cc{\cs} \geq 0, \r \geq 0, \sum_{i=1}^{\cs} \cc{i} +\r \leq \nslot \Big\}.
\end{align}

The following theorem establishes a recursion for the state generating function.
\begin{theorem}\label{theorem_state_gen_fun}
  Consider a contention period with $\nslot$ slots of $\cs$ different types and $\nuser$ contending users, which in a slot of type $h$ are active with probability $\paccess_{h}$. For $u=n, n-1,\hdots,1$, we have
  \begin{align}\label{eq:theorem_new}
  & \Pu (\xc{1},\xc{2},\hdots, \xc{\cs},\y) = \\
   &\frac{1}{\y} \Bigg[ \Pu \Bigg(  \xc{1}(1-\puc{1}) + \y \puc{1}, \xc{2}(1-\puc{2}) + \y \puc{2},   
   \hdots, \xc{\cs}(1-\puc{\cs}) + \y \puc{\cs}, \frac{1}{u} + \y \left( 1 -\frac{1}{u} \right)\Bigg)  \\
   &  - \Pu \left( \xc{1} (1-\puc{1}),\xc{2} (1-\puc{2}), \hdots, \xc{\cs} (1-\puc{\cs}), \frac{1}{u}\right)\Bigg]
  \end{align}
  where  $\puc{h}$ is given by
  \[
\puc{h} =  \frac{ \mathlarger {\sum}\limits_{d=2}^{\nuser-u+2}   \Omega_{h,d} \, d (d-1)\frac{1}{\nuser}  \frac{u-1}{\nuser-1}  \frac{\binom{\nuser-u}{d-2}}{\binom{\nuser-2}{d-2}} }
{  1 -  \mathlarger{\sum}\limits_{d=1}^{\myop{\nuser-u+1}}    \Omega_{h,d} \,u \frac{\binom{\nuser-u}{d-1}}{\binom{\nuser}{d}}
-  \mathlarger{\sum}\limits_{d=0}^{\myop{\nuser-u}}   \Omega_{h,d} \frac{\binom{\nuser-u}{d}}{\binom{\nuser}{d}}    }
\]
and initial condition given by
\begin{equation}
\P_{\nuser} (\xc{1},\xc{2},\hdots, \xc{\cs},\y)  = 
\mkern-2mu \frac{1}{y}  \prod_{h=1}^{\cs} \big( (1-\Omega_{h,1}-\Omega_{h,0}) \x + \Omega_{h,1} \y + \Omega_{h,0} \big)^m
\mkern-2mu
-  \big( (1-\Omega_{h,1} - \Omega_{h,0}) \x + \Omega_{h,0} \big)^m .  \label{eq:p_k}
\end{equation}

\end{theorem}
\begin{IEEEproof}
See Appendix~\ref{sec:theorem_state_proof}.
\end{IEEEproof}

Theorem~\ref{theorem_state_gen_fun} can be used to derive the probability of decoding failure in a similar way as done for the LT decoder in \cite{shokrollahi2009theoryraptor}. In particular, we have that the probability that the decoder fails when exactly $u$ users are undecoded is $1-\Pu(\mathbf{1})$, where $\mathbf{1}$ is the all one vector. This corresponds to the probability that the ripple is empty. Hence, the probability that exactly $u$ users remain unresolved after a contention period of $\nslot$ slots, ${\per}_u$, corresponds to:
\[
{\per}_u = 1-\Pu(\mathbf{1}).
\]

\subsection{First Moment}\label{sec:diff_eqs_first}

Denote by $C_h(u)$ and $R(u)$ the expected number of slots in the $h$-th cloud and the ripple, respectively. We have
\begin{equation}\label{eq:c_u_h_def}
C_h(u) := \sum_{ \validstates} \cc{h} \, \pcccru 
\end{equation}
and
\begin{equation}\label{eq:r_u_def}
R(u) : = \sum_{ \validstates }  (\r-1) \, \pcccru 
\end{equation}
where we recall that $\validstates$ is the set of valid decoder states given in \eqref{eq:valid_states}.

It is possible to express $C_h(u)$ and $R(u)$ as the first-order derivatives of the state generating function \eqref{eq:theorem_new} in Theorem~\ref{theorem_state_gen_fun}, evaluated at $\mathbf{1}$, where $\mathbf{1}$ is the all one vector. In particular, we have
\begin{align}\label{eq:c_u_h_def_2}
C_h(u) &:= 
\frac{\partial}{\partial\xc{h}} \Pu (\xc{1},\xc{2},\hdots, \xc{\cs},\y)|_{\mathbf{1}}
\end{align}
and
\begin{align}\label{eq:r_u_def_2}
R(u) &: = \frac{\partial}{\partial \y} \Pu (\xc{1},\xc{2},\hdots, \xc{\cs},\y)|_{\mathbf{1}}.
\end{align}

Let us first focus on $C_h(u)$. From \eqref{eq:c_u_h_def_2} we have
\begin{align}\label{eq:c_h_der}
C_h(u-1) &= (1-\puc{h}) C_h(u) 
- (1-\puc{h}) \frac{\partial}{\partial \xc{h}} \Pu \left( 1-\puc{1}, \hdots, 1-\puc{k}, \frac{1}{u} \right)
\end{align}
In \cite{Maatouk:phd} it was shown that, for $r\geq 6$, the drift term \eqref{eq:c_h_der} is $\mathcal{O}(1/\nuser^2)$, leading to
\begin{align}\label{eq:c_h_der2}
C_h(u-1) &= (1-\puc{h}) C_h(u) + \mathcal{O}(1/\nuser^2).
\end{align}

In a similar way, the expression for $R(u)$ can be obtained differentiating both sides of the recursion in \eqref{eq:theorem_new} with respect to $\y$ and evaluating the expression at $\mathbf{1}$,
\begin{align}\label{eq:r_der}
R(u-1) &= \left( 1 - \frac{1}{u} \right) R(u)+ \sum_{h=1}^{\cs} \puc{h} C_h(u) - \Pu( \mathbf{1}) \\
&+  \Pu(1-\puc{1},1-\puc{2},\hdots,1-\puc{\cs}, 1/u).
\end{align}
We again make use of a result in \cite{Maatouk:phd}, where it was shown that for $\r \geq 5$ the residual term
\[
- \Pu( \mathbf{1}) +  \Pu(1-\puc{1},1-\puc{2},\hdots,1-\puc{\cs}, 1/u)
\]
can be approximated as $-1+ \mathcal{O}(1/\nuser^2)$.

Recall that in frameless ALOHA, due to the contention mechanism, in a slot of type $h$, users are active with a probability $\paccessc{h}$. This induces a binomial slot degree distribution given by eq.~\eqref{eq:omega_h_j}.
If for a fixed contention period length, we look at the contention graph of frameless ALOHA from the perspective of slots, this is equivalent to saying that slots choose their neighbours uniformly at random and \emph{without replacement} (a user can not be active multiple times in a slot).

Following \cite{Karp2004}, \cite{Maatouk:2012} the frameless ALOHA decoder can be approximated by introducing the assumption that slots nodes choose their neighbors \emph{with replacement} in the bipartite graph representation of the contention process, i.e., the same user node can be chosen several times in the same slot. If a slot node has an odd number of edges connected to a user node, the user node will be active in that slot. If the number of edges is even $\{0,2,\hdots \}$, the user is not active in that slot. This approximation results in a slight decrease of the slot access probability $\paccessc{h}$. Nevertheless, intuitively the approximation becomes tighter as the number of users $\nuser$ increases, since it becomes less and less likely that a slot chooses several times the same user.

Under the replacement assumption, the expression of $\puc{h}$ becomes:
\begin{equation}\label{eq:puc_x}
\puc{h} = \frac{1}{\nuser} f_{h}\left(\frac{u}{\nuser}\right) - \frac{1}{\nuser^2} g_h \left( \frac{u}{\nuser} \right)= \frac{1}{\nuser} f_h \left(\frac{u}{\nuser}\right) + \mathcal{O}(1/\nuser^2)
\end{equation}
where
\[
f_h(x) = \frac{x \Omega_h''(1-x)}{1- x \, \Omega_h'(1-x) - \Omega_h(1-x)},
\]
\[
g_h(x) = \frac{f_h(x)}{x},
\]
and $\Omega_h(x)$ is the generator polynomial of the slot degree distribution of slots of type $h$,
\[
\Omega_h(x) = \sum_d \Omega_{h,d} \, x^d.
\]

From \eqref{eq:c_h_der2} and \eqref{eq:puc_x} we obtain the following difference equation for the $h$-th cloud
\begin{align}\label{eq:c_h_diff}
C_h(u) - C_h(u-1) = f_{h}\left(\frac{u}{\nuser}\right)  C_h(u) + \mathcal{O}(1/\nuser^2)
\end{align}
Similarly, from \eqref{eq:r_der} and \eqref{eq:puc_x} we obtain
\begin{align}
R(u) - R(u-1) & = \frac{1}{u} R(u) - \sum_{h=1}^{\cs} f_{h}\left(\frac{u}{\nuser}\right)  C_h(u) +1 
+ \mathcal{O}(1/\nuser^2) \label{eq:r_diff}
\end{align}

Let us now define the expected normalized size of the clouds and ripple respectively as
\[
C_h(\xi):= C_h(u)/\nslot
\] and
\[
R(\xi) := R(u)/ \nslot.
\]
Making use of these definitions, and assuming $\nslot = (1+\epsilon) \nuser$, we can divide both sides of \eqref{eq:c_h_diff} by $\nslot$ to obtain
\begin{align}\label{eq:c_h_diff_xi}
C_h(\xi) - C_h(\xi-1/\nuser) =  f_{h}(\xi)  C_h(\xi) + \mathcal{O}(1/\nuser^3)
\end{align}
and the same can be done for \eqref{eq:r_diff} leading to
\begin{align}\label{eq:r_diff_xi}
R(\xi) - R(\xi-1/\nuser) &= \frac{1}{\xi \nuser}  R(\xi) - \sum_{h=1}^{\cs} f_{h}(\xi)  C_h(\xi) 
 + \frac{1}{\nuser(1+\epsilon)} +\mathcal{O}(1/\nuser^3).
\end{align}

As shown in \cite{Maatouk:phd}, it is possible to approximate $C_h(\xi)$ $R(\xi)$ respectively by $\hat C_h(\xi)$ $\hat R(\xi)$, which are the solutions to the following differential equations
\begin{align}\label{eq:c_h_diff_x}
\hat C_h'(\x) = f_{h}(\x)  \hat C_h(\x)
\end{align}
and
\begin{align}\label{eq:r_diff_xx}
\hat R'(\x) = \frac{\hat R(\x)}{\x}   - \sum_{h=1}^{\cs} f_{h}(\x)  \hat C_h(\x) + \frac{1}{1+\epsilon}.
\end{align}
These approximations are tight during almost all the decoding process except for the last few users that are decoded. In particular, it was shown in \cite{Maatouk:phd} that as long as $u$ is a constant fraction of $\nuser$, we have
\begin{equation}\label{eq:cloud_approx_error}
C_h(u) / \nslot = \hat C_h (u/k) +  \mathcal{O}(1/\nslot)
\end{equation}
and
\begin{equation}\label{eq:ripple_approx_error}
R(u) / \nslot = \hat R (u/k) + \mathcal{O}(1/\nslot).
\end{equation}
Thus, the approximations become tighter for increasing $\nslot$.

Finally, the expression for $\hat C_h(\x)$ and $\hat R(\x)$ are obtained by solving the differential equations \eqref{eq:c_h_diff_x} and \eqref{eq:r_diff_xx}, giving rise to the following solutions \cite{Maatouk:2012}
\[
\hat C_h(x) = \cho \big( 1- x \, \Omega_h'(1-x) - \Omega_h(1-x) \big)
\]
\[
\hat R(x) = x \left( \sum_{h=1}^{\cs} \cho \Omega_h'(1-x) +  \frac{\nuser}{\nslot} \log(x) + \ro  \right)
\]
where the values of the parameters $\cho$ and $\ro$ are determined by the initial conditions to the differential equations. In particular, for the clouds we have
\begin{align}
C_h(\nuser) &= \sum_{\validstates} \cc{h} \, \pcccrn = \nslotc{h} \left( 1 - \Omega_{h,0} - \Omega_{h,1} \right)  \left( 1 - \prod_{l=1}^{\cs} \big(1-\Omega_{l,1} \big)^{\nslotc{h}}\right).
\end{align}
Hence, by imposing $\hat C_h(x=1) = C_h(\nuser) / \nslot$, we obtain
\begin{equation}\label{eq:ch0}
\cho = \frac{\nslotc{h}}{\nslot} \left( 1 - \prod_{l=1}^{\cs} \big(1-\Omega_{l,1} \big)^{\nslotc{l}}\right).
\end{equation}
For the ripple we have
\begin{align}
R(\nuser) &=  \sum_{\validstates} (\r-1) \, \pcccrn 
= \sum_{h=1}^{\cs} \nslotc{h}\Omega_{h,1} - 1 +  \prod_{h=1}^{\cs} \big( 1- \Omega_{h,1}\big)^{\nslotc{h}} .
\end{align}
Thus, imposing $\hat R(x=1) = R(\nuser)/ \nslot$ yields
\begin{align}\label{eq:r0}
\ro &= \sum_{h=1}^{\cs}   \frac{\nslotc{h}}{\nslot} \Omega_{h,1} \prod_{h=1}^{\cs}  \big( 1-\Omega_{h,1}  \big)^{\nslotc{h}} 
 - \frac{1}{\nslot} \left( 1- \prod_{h=1}^{\cs} \big( 1-\Omega_{h,1}  \big)^{\nslotc{h}}\right) .
\end{align}

\subsection{Second Moment}\label{sec:diff_eqs_second}

In the following, we shall focus on the variance of the ripple, $\sigma_R(u)$. By definition, we have
\[
\sigma_R(u) =  \sum_{ \validstates} (r-1)^2 \pcccru - R(u)^2.
\]

Experimentally, we made the observation that the distribution of the ripple $\Ripp_u$ is approximately binomial (see Fig.~\ref{fig:ripple_dist}). Under the assumption that $\Ripp_u$ is binomially distributed with parameters $\nslot$ and $\rho=R(u)/\nslot$, a continuous approximation of the variance of the normalized ripple $\sigma_R^2(u)/m^2$ is given by
\[
\hat \sigma_R^2(x) = \frac{\rho (1-\rho)\nslot}{\nslot^2} = \frac{\rho (1-\rho)}{\nslot} .
\]

\subsection{Numerical Results}\label{sec:diff_eqs_num}

\begin{figure}[t]
\subfloat[Normalized expected cloud]{
\includegraphics[width=0.48\columnwidth]{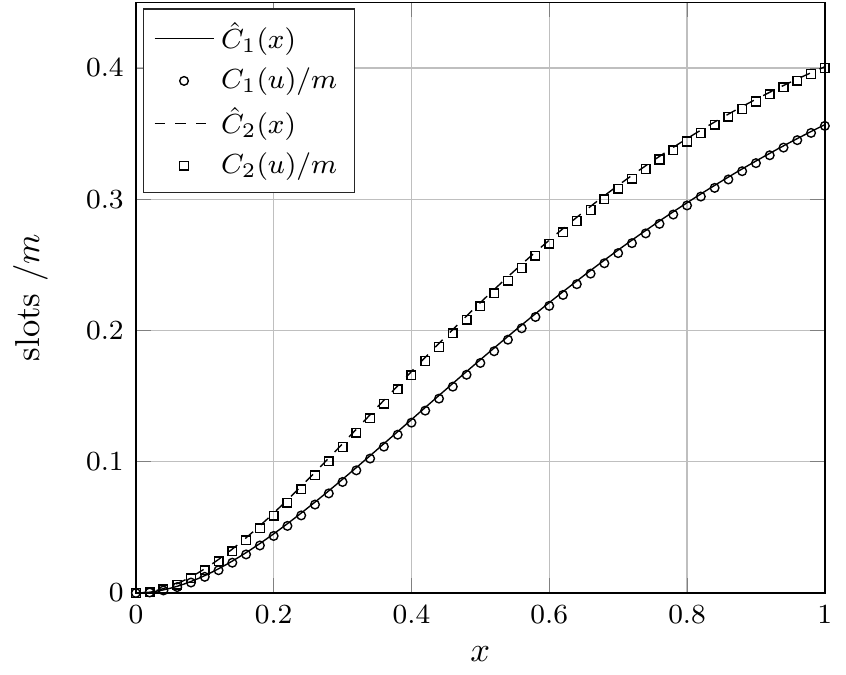}
\label{fig_a}
}
\subfloat[Normalized expected ripple]{
\includegraphics[width=0.48\columnwidth]{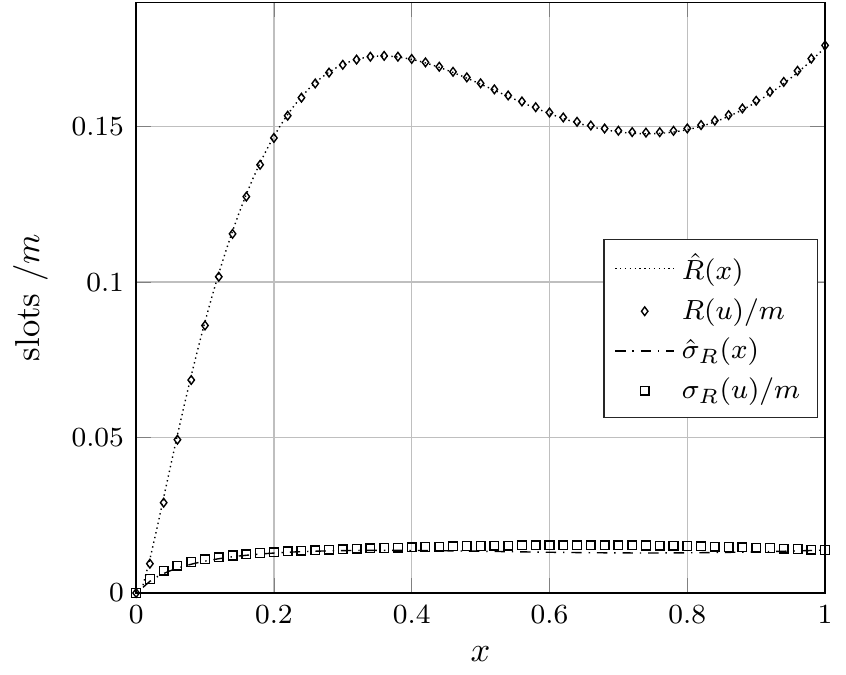}
\label{fig_b}
}
\caption{Normalized expected ripple and clouds, as well as the normalized standard deviation of the ripple and their continuous approximations for $\nuser=400$ users, $k=2$ slot types, $\nslotc{1} = \nslotc{2} = 380$ slots, $\betac{1}=2.5$ and $\betac{2}=3$. }
\label{fig:ripple_cloud_mean_std}
\end{figure}

Fig.~\ref{fig:ripple_cloud_mean_std} shows the expected normalized ripple and clouds sizes, as well as the normalized standard deviation of the ripple and their continuous approximation. The setting considered is $\nuser= 400$ users, $k=2$ slot types, $\nslotc{1} = \nslotc{2} = 380$ slots, $\betac{1}=2.5$ and $\betac{2}=3$.
We can observe how the continuous approximation $\hat R(x)$ is very close to the actual expected normalized ripple size $R(u)/\nslot$. There exists also a tight match between the  normalized clouds $C_1(u)/\nslot$ and $C_2(u)/\nslot$ and their continuous approximations $\hat C_1(c)$ and $\hat C_2(x)$. We can also observe how the match between $\sigma_R(u)/m$ and $\hat \sigma_R(x)$ is tight.

\begin{figure}[t]
\subfloat[$R$]{
\includegraphics[width=0.45\columnwidth]{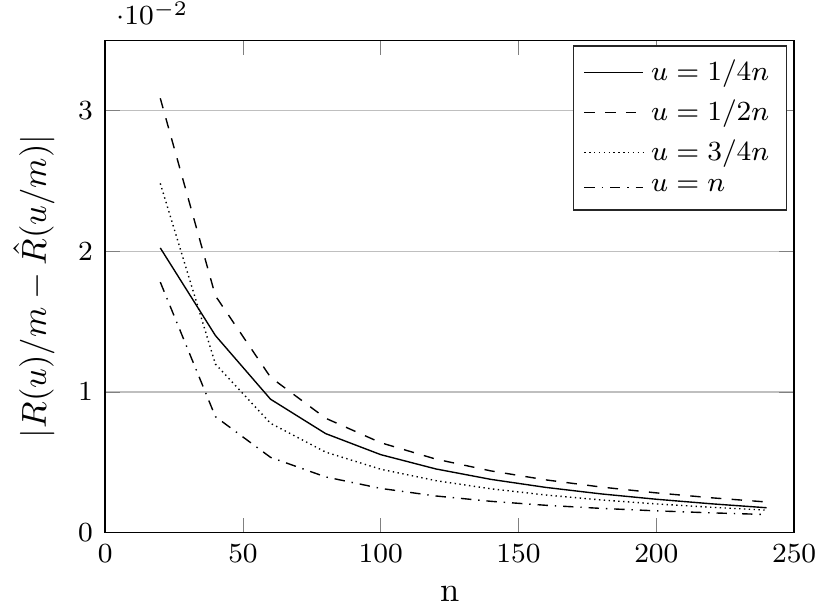}
\label{fig_R_scaling}
}
\subfloat[$\sigma_R^2$]{
\includegraphics[width=0.45\columnwidth]{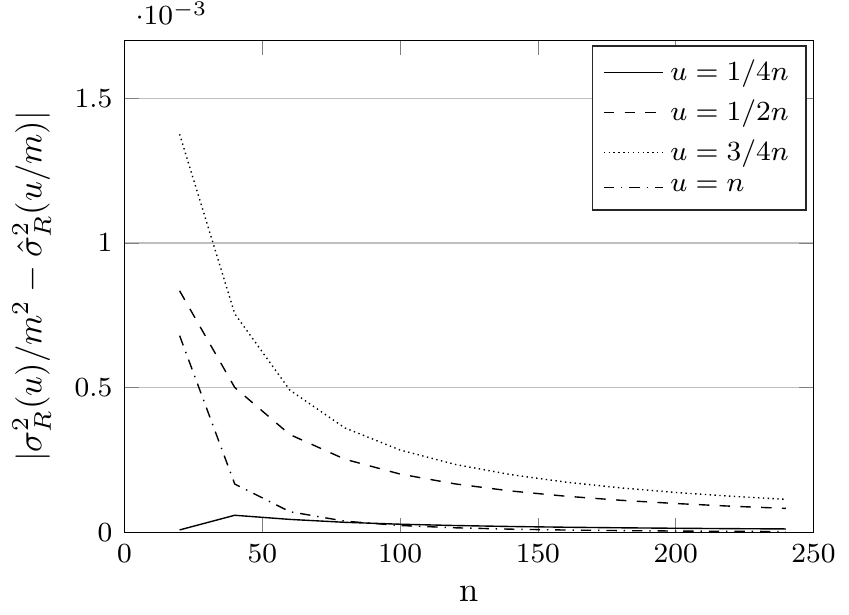}
\label{fig_R_var_scaling}
}

\subfloat[$C_1$]{
\includegraphics[width=0.45\columnwidth]{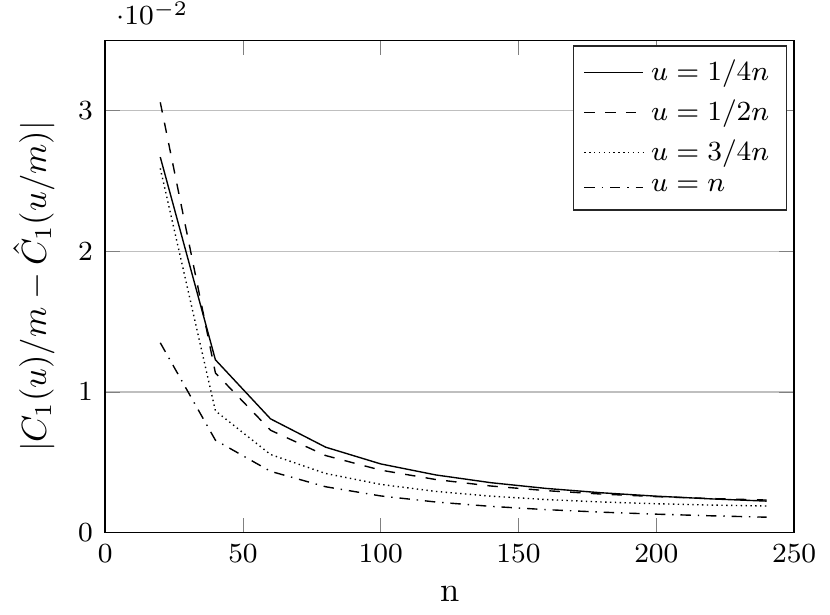}
\label{fig_C_1_scaling}
}
\subfloat[$C_2$]{
\includegraphics[width=0.45\columnwidth]{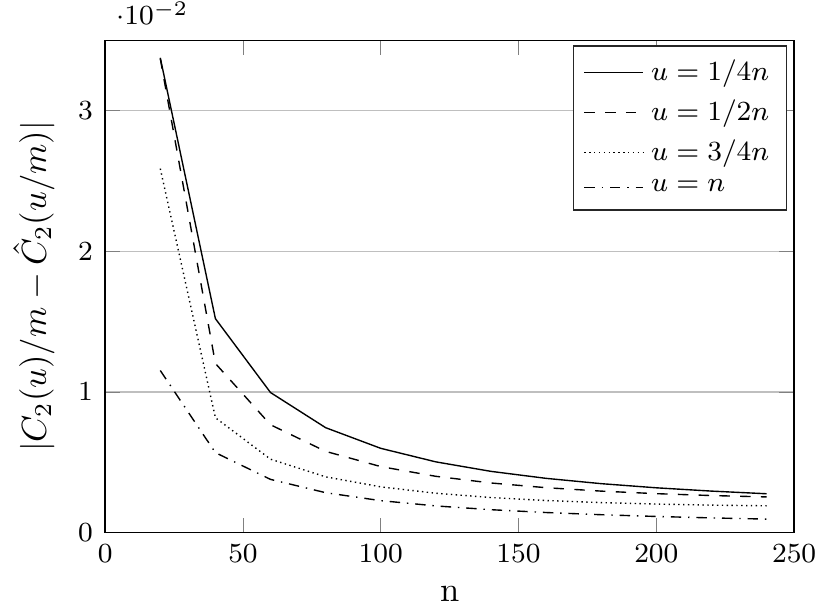}
\label{fig_C_2_scaling}
}
\caption{Absolute approximation error for $k=2$ slot types, $\betac{1}=2.5$ and $\betac{2}=3$, ${\nslotc{1}/\nuser = \nslotc{2}/\nuser = 19 / \nuser}$
as a function of $\nuser$. The different curves represent different decoding instants ($u={\nuser/4, \nuser/2, 3 \nuser/4, \nuser }$). Subfigures (a), (b), (c) and (d) represent the approximation error for $R$, $\sigma_R^2$, $C_1$ and $C_2$, respectively.}
\label{fig:approx_error_scaling}
\end{figure}

Fig.~\ref{fig:approx_error_scaling} shows the absolute approximation error for $R$, $\sigma_R^2$, $C_1$ and $C_2$ as a function of $\nuser$ for a setting with $k=2$ slot types, $\betac{1}=2.5$ and $\betac{2}=3$, ${\nslotc{1}/\nuser = \nslotc{2}/\nuser = 19 / \nuser}$. In particular, every subfigure shows the error at 4 different decoding instants ($u={\nuser/4, \nuser/2, 3 \nuser/4, \nuser }$). All curves in the figure were obtained applying Theorem~\ref{theorem:class}. We can observe how in all cases the approximation error decreases as the number of users $\nuser$ increases. The approximation error for the expected ripple and cloud sizes decreases approximately as $1/\nuser$, which is in line with the error terms in  \eqref{eq:cloud_approx_error} and \eqref{eq:ripple_approx_error}. It is also remarkable how the approximation error for the ripple variance $\sigma_R^2$ also decreases as the number of users $\nuser$ increases, although this approximation was heuristically derived.

\subsection{Approximation to the PER}\label{sec:diff_eqs_per}

We shall now see how the continuous approximations derived in this section can be used in order to estimate the expected packet error rate, $\per$, i.e., the probability that a user is not resolved. In particular, we shall make use of the continuous approximations of  the expected ripple $\hat R(x)$ and of its standard deviation $\hat \sigma_R(x)$.

\begin{figure}[t]
\centering
\subfloat[u=n=400]{
\includegraphics[width=0.41\columnwidth]{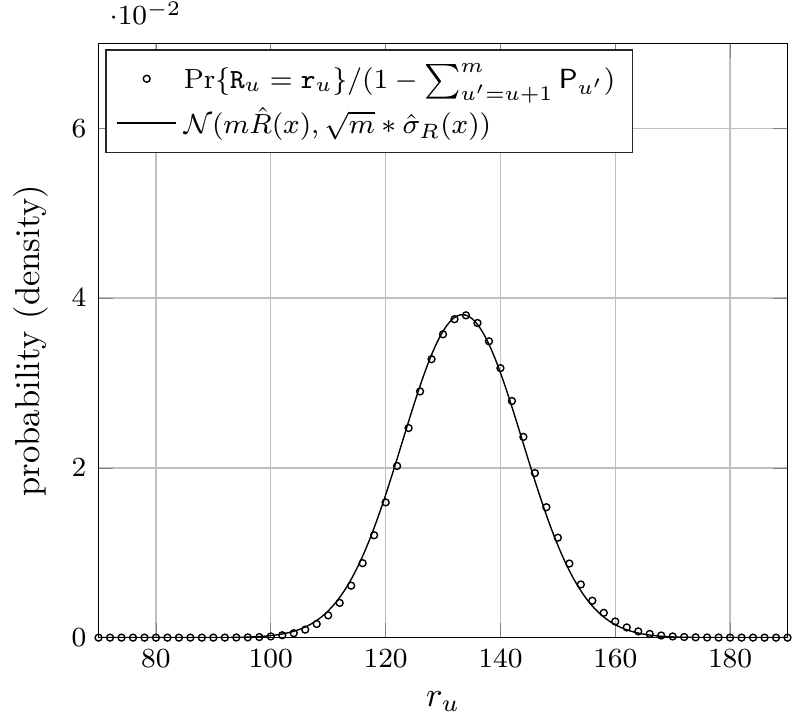}
\label{fig_first_case}
}
\subfloat[u=300]{
\includegraphics[width=0.41\columnwidth]{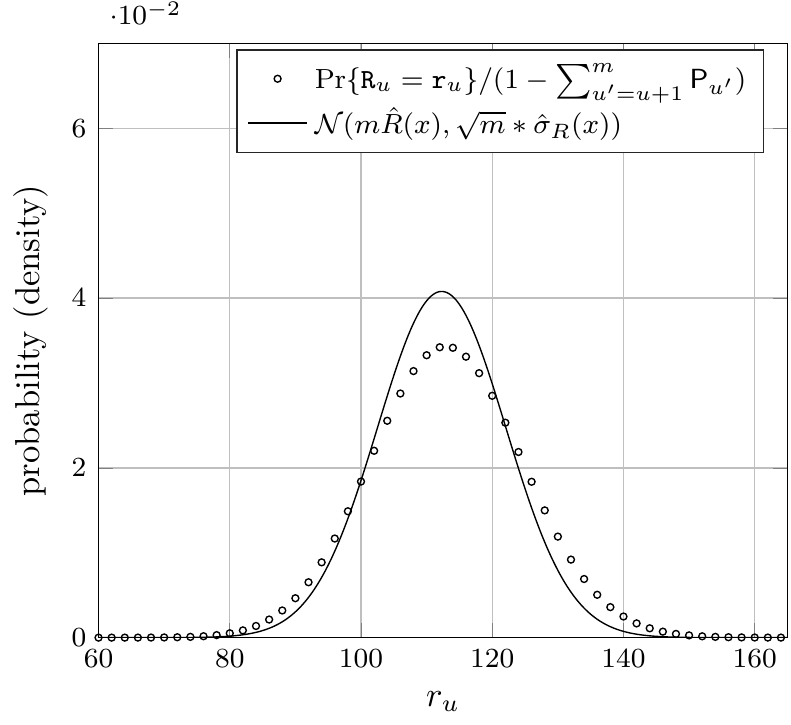}
\label{fig_second_case}
}

\subfloat[u=200]{
\includegraphics[width=0.41\columnwidth]{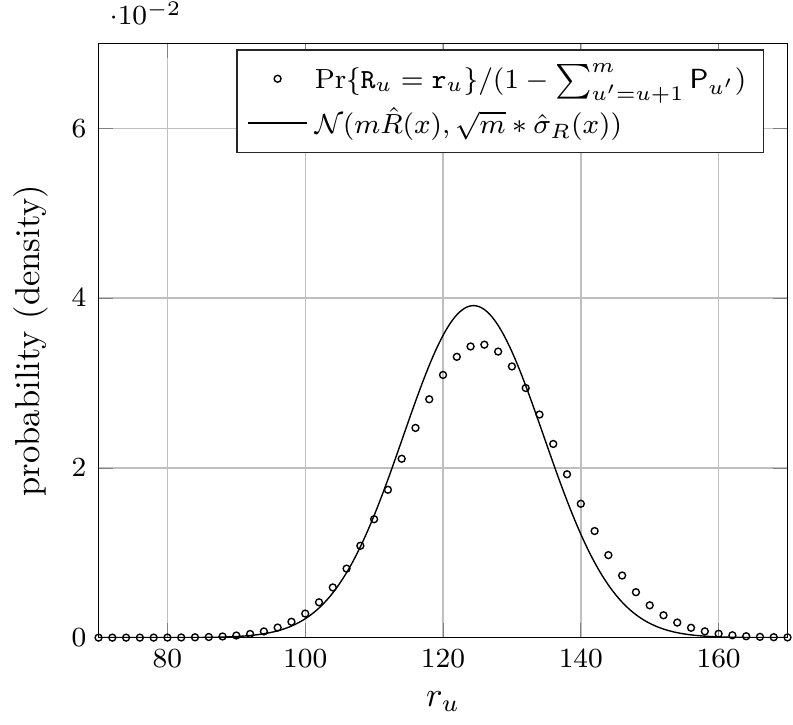}
\label{fig_third_case}
}
\subfloat[u=100]{
\includegraphics[width=0.41\columnwidth]{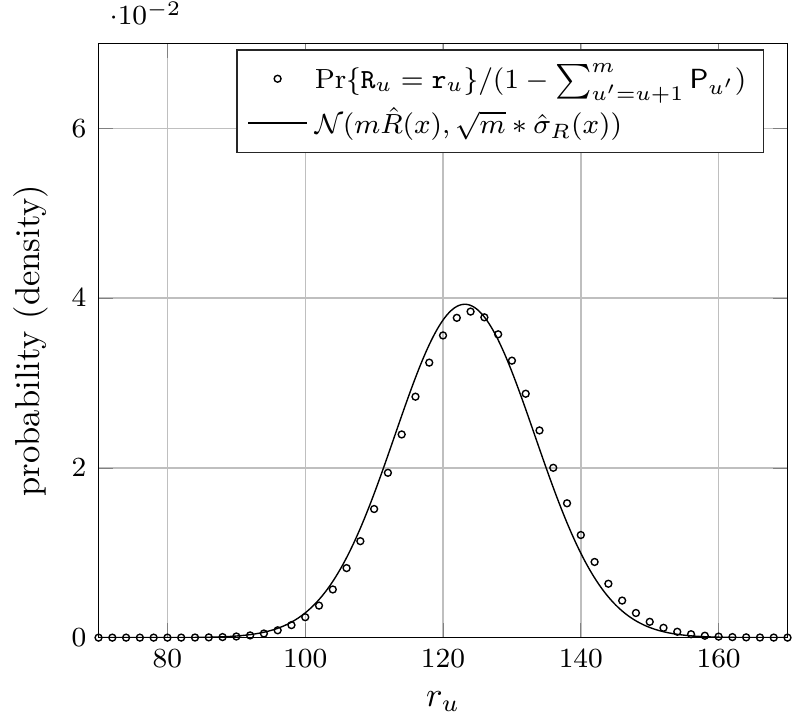}
\label{fig_fourth_case}
}
\caption{Probability mass function of the ripple cardinality for $\nuser= 400$ users, $k=2$ slot types, $\nslotc{1} = \nslotc{2} = 380$ slots, $\betac{1}=2.5$ and $\betac{2}=3$ at different decoding instants. The solid line shows a Gaussian distribution with mean $\nslot \hat R(x)$ and standard deviation $\sqrt \nslot * \hat \sigma_R(x))$.}
\label{fig:ripple_dist}
\end{figure}

Fig.~\ref{fig:ripple_dist} shows the distribution of the ripple for $\nuser= 400$ users, $\nslotc{1} = \nslotc{2} = 380$, $\betac{1}=2.5$ and $\betac{2}=3$ at 4 different instants during the decoding process. In particular, the figure shows the probability mass function of the ripple cardinality when $u$ users are undecoded conditioned to the decoding process not failing for $u' >u$. That is,
\[
\frac{\Pr\{\Ru=\ru\}} { 1 - \sum_{u'=u+1}^{\nslot} {\per}_{u'} } = \frac{\Pr\{\Ru=\ru\}} { \sum_{\ru=0}^{\nslot} \Pr\{\Ru = \ru \} }.
\]
Moreover, the figure also shows the probability density function of Gaussian random variables with  mean $\nslot \hat R(x)$ and standard deviation $\sqrt \nslot * \hat \sigma_R(x)$. We make several observations in relation to this figure. The first is that the ripple cardinality has a bell shaped distribution across the whole decoding process, which resembles a Gaussian distribution. The second is that a Gaussian curve with  mean $\nslot \hat R(x)$ and standard deviation $\sqrt \nslot * \hat \sigma_R(x)$ is reasonably close to the actual distribution of the ripple. 
In particular, we can observe how the Gaussian approximation is tight at the beginning and the end of the decoding process, but it is not as tight in the middle of the decoding process, owing to the aforementioned overestimation of the standard deviation of the ripple size. We would also like to remark that, while random variable $\Ru$ is discrete, we are approximating it by a continuous (Gaussian) random variable.

Based on this observation, we can approximate the probability that exactly $u$ users remain unresolved after the end of the contention period,  ${\per}_u$, as
\begin{equation}\label{eq:per_u_approx}
\hat \per_u = \mathtt{Q} \left( \frac{\hat R(u/\nslot) } { \hat \sigma_R(u/\nslot) } \right) \left(  1- \sum_{l=u+1}^{\nuser} \hat \per_l \right)
\end{equation}
for $u< \nuser$ and
\begin{equation}\label{eq:PMF_ru_norm}
\hat \per_\nuser =  \mathtt{Q} \left( \frac{\hat R(\nuser/\nslot)} { \hat \sigma_R(\nuser/\nslot) }\right)
\end{equation}
for $u = \nuser$,
where $\mathtt{Q}$ is the tail distribution function of the standard normal distribution. The first term in \eqref{eq:per_u_approx} corresponds to the probability that a Gaussian distribution with mean $\hat R(u/\nslot)$ and standard deviation  $\hat \sigma_R(u/\nslot)$ takes a negative value. Hence, this first term corresponds to the probability that our estimation of the ripple is negative. The second term in \eqref{eq:per_u_approx} implies we are conditioning to the decoding process in the previous decoding stages ($\nuser$ down to $u + 1$) being successful. An alternative way of interpreting this is that, according to our definition of ${\per}_u$, we have
$\sum_{u=0}^{\nuser} {\per}_u =1$
and by introducing the second term we are enforcing
$\sum_{u=0}^{\nuser} \hat {\per}_u =1$.
Thus, we can approximate the expected packet error rate $\per$ as
\begin{equation}\label{eq:hatper}
\hat \per = \sum_{u=0}^{\nuser} u /\nuser \, \hat  \per_u .
\end{equation}
\begin{figure}[t]
        \centering
        \includegraphics[width=0.50\columnwidth]{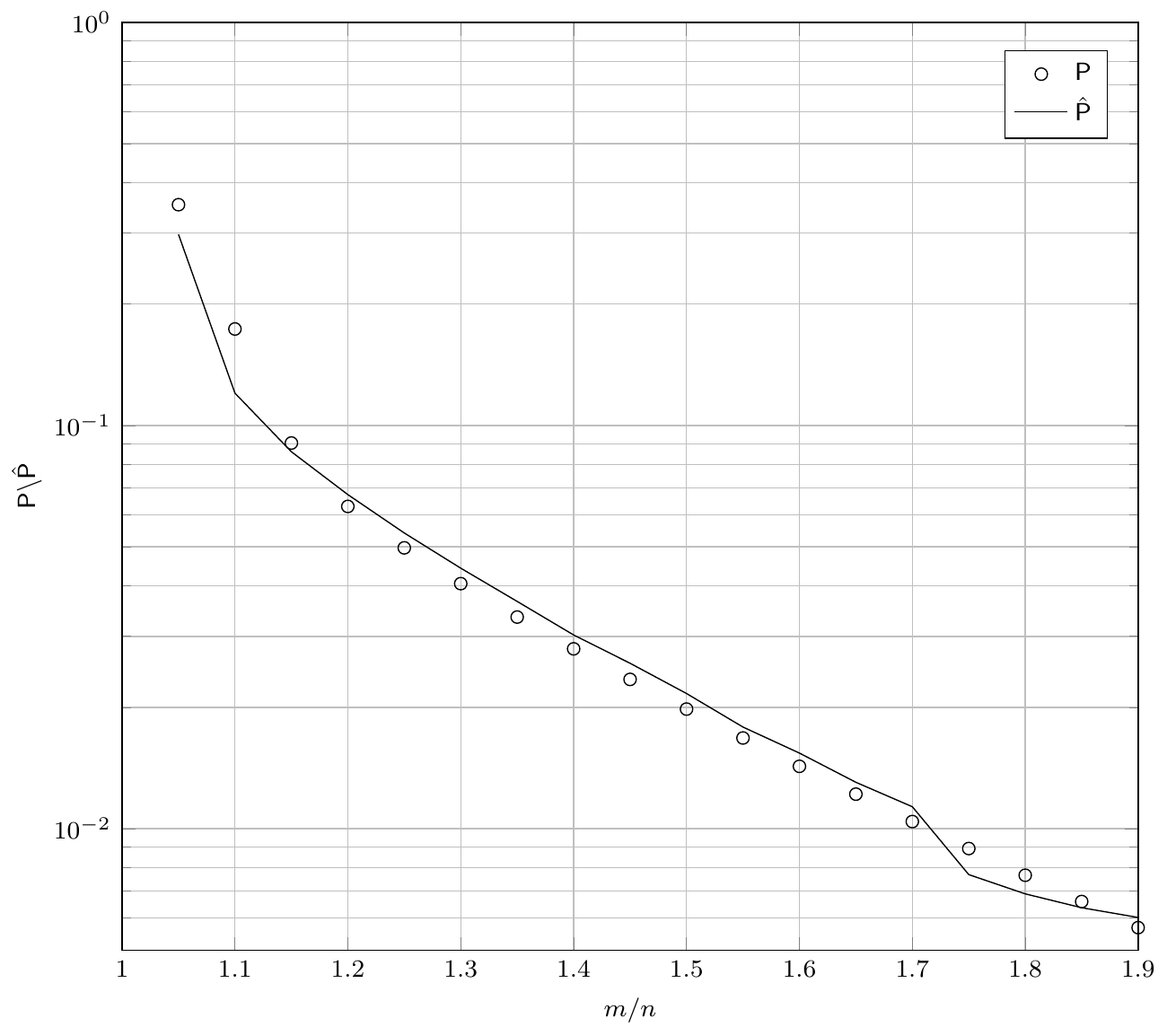}
        \caption{Packet error rate $\per$ and its approximation $\hat \per$ as a function of $\nslot/\nuser$ for $\nuser= 400$ users, $k=2$ slot types, $\nslotc{1} = \nslotc{2} = \nslot/2$, $\betac{1}=2.5$ and $\betac{2}=3$. }
        \label{fig:per_approx}
\end{figure}

Fig.~\ref{fig:per_approx} show the packet error rate $\per$ as well as its approximation $\hat \per$ as a function of $\nslot/\nuser$, for $\nuser= 400$ users, $k=2$ slot types, $\nslotc{1} = \nslotc{2} = \nslot/2$, $\betac{1}=2.5$ and $\betac{2}=3$. The packet error rate $\per$ has been obtained using \eqref{eq:per} and applying Theorem~\ref{theorem:class}, which is exact, whereas the results for $\hat \per$ have been obtained using \eqref{eq:hatper}.
We can observe how the approximation of the packet error rate is tight. The fact that the estimation of the packet error rate is tight despite the fact that the standard deviation of the ripple is overestimated in the middle of the decoding process can be attributed to the fact that the dominating error event is that decoding stops at the end of the decoding process (when almost all users have been recovered). As we saw in Fig.~\ref{fig:ripple_dist}, the estimate of the standard deviation of the ripple in this regime (at the end of the decoding process) is tight.


\section{Dynamic Feedback}\label{sec:dynamic}
\begin{figure}[t]
	\centering
	\includegraphics[width=0.99\columnwidth]{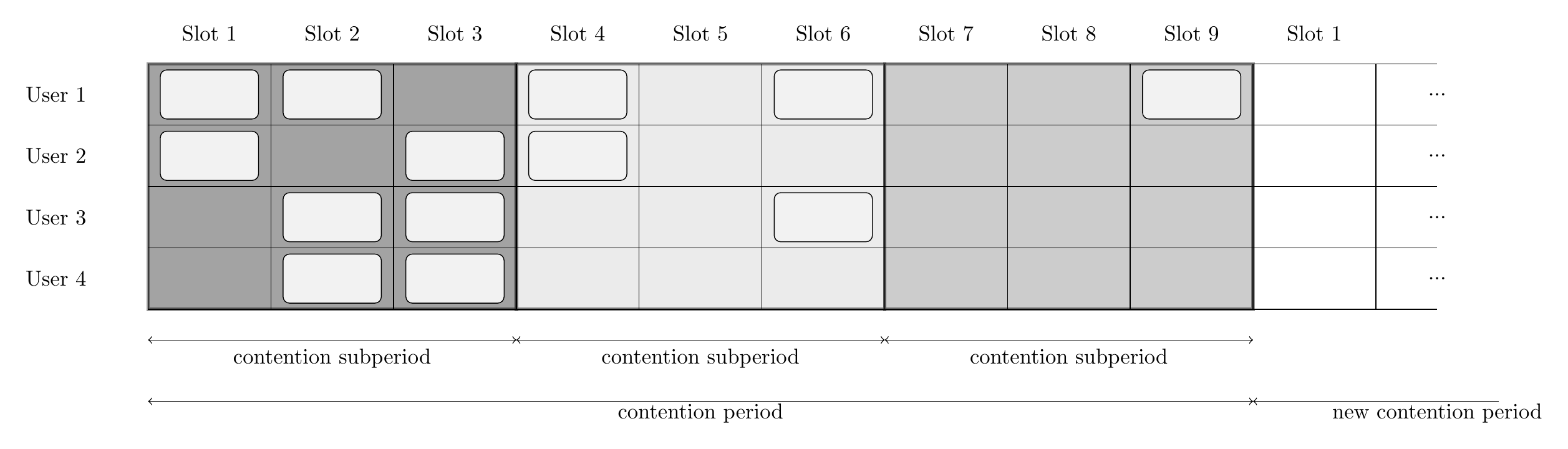}
	\caption{An example of contention in frameless ALOHA with dynamic feedback with four users, where the contention subperiods have length $\tp=3$ slots.  In this example the slot access probability for the first contention subperiod,  $\paccessc{1}$, happens to be set too high. As a consequence, all the slots in the first contention subperiod present collisions. After the first $\tp=3$ slots the \ac{AP} notifies the users to use $\paccessc{2}$ in the next contention subperiod, which spans slots 4, 5 and 6. However, after the second contention subperiod still no users can be decoded. The \ac{AP} then notifies the users of the slot access probability to be used in the next contention subperiod, $\paccessc{3}$. Finally, in the third contention subperiod a singleton slot is received, which allows the \ac{AP} to decode all users leveraging on \ac{SIC}. The \ac{AP} notifies all users that a new contention period starts.}
	\label{fig:frameless_adaptive}
\end{figure}

In this section show how the approximation of the analysis presented in this paper can be used together with feedback in order to improve the performance of frameless ALOHA.\footnote{The idea of using feedback in frameless ALOHA to adapt the slot access probability has also been explored in \cite{Ogata:2019}, where the slot access probability is adapted after every slot in order to keep the expected number of unresolved \emph{active} users per slot constant. However, the setting considered in \cite{Ogata:2019} is different to the one considered in this paper, since it assumes that collisions of size two can be detected and resolved by employing ZigZag decoding.}
We assume that a perfect and zero-delay feedback channel exists from the access point to the terminals. This channel is used to notify users of the value of the slot access probability which has to be used in upcoming slots. 
We also assume that the contention period is divided into subperiods of duration $\tp$ slots,  as depicted in Fig.~\ref{fig:frameless_adaptive}. All slots belonging to the same subperiod are of the same type, i.e., they are characterized by the same slot access probability. 
At the end of the $\he$-th subperiod (after $\he \cdot \tp$ slots have elapsed), the \ac{AP} runs the decoding algorithm to decode as many users as possible (until the ripple is empty).
At this stage, denote by $u$ the number of users which are yet unresolved, and by $\S{u}$ the decoder state, with
\[
\S{u}:=( \cuc{1}, \cuc{2}, \cdots, \cuc{\he}, \ru=0 )
\]
where $\cuc{i}$ is the number of slots in the $i$-th cloud.
Next, based on the number of unresolved users $u$, the decoder state $\S{u}$, and using as latency target a contention period duration of $\nslotT$ slots, the decoder computes the slot access probability to be used in the next subperiod, $\paccessc{\he+1}$, and signals its to the users.

In particular, in order to determine the value of $\paccessc{\he+1}$ to be used, the \ac{AP} assumes that the remaining slots up to $\nslotT$ will be of class $\he+1$. Thus, it sets
$\nslotc{\he+1}= \nslotT - \he \, \tp$.
The \ac{AP} now needs to determine the initial state of the frameless ALOHA decoder, given $\nslotc{\he+1}$, $u$ and $\S{u}$. The first step is determining the
slot degree distribution of the slots of class $\he+1$, $\Omega_{\he+1,j}'$, taking into account that $u$ out of $\nuser$ users have already been decoded.  Let us denote by $\St$ the random variable associated to the number of users which are active in a slot and by $\st$ its realization. Let us also denote by $\J$ the random variable associated to the number of \emph{unresolved} users which are active in a slot and by $\j$ its realization. We have
\begin{align} \label{eq:de_dist_virtual}
\Omega_{\he+1,\j}' &= \Pr\{\J=\j\}= \sum_{\st=j}^{\nusernew} \Pr\{\St=\st\}  \Pr\{\J=\j| \St=\st\}  \\
&= \sum_{\st=\j}^{\nusernew} { \nuser \choose \st } \left(\paccessc{\he+1} \right)^\st \left( 1 - \paccessc{\he+1} \right)^{\nuser - \st} 
                                       { \st \choose \j } \left(\frac{\nusernew}{\nuser} \right)^\j  \left( 1 - \frac{\nusernew}{\nuser} \right)^{\st- \j}. 
\end{align}
Finally, the initial state distribution of the frameless ALOHA decoder, $\S{u}'$, is given by
\begin{align}
\Pr  \{ \S{u}' =   \cuc{1}, \cuc{2}, \cdots, \cuc{\he+1}, \ru \} &=
 \frac{\nslotc{\he+1} !}{\c_{\he+1,\nusernew}! \, \r_{\nusernew}!\, (\nslotc{\he+1}-\c_{\he+1,\nuser}-\r_{\nusernew})!}  \\
 & \mkern-20mu \times  \left( 1-\Omega_{\he+1,1}- \Omega_{\he+1,0}\right)^{\c_{\he+1,\nusernew}}\, {\Omega_{\he+1,1}}^{\r_{\nusernew}} \, {\Omega_{\he+1,0}}^{\nslotc{\he+1} -\c_{\he+1,\nusernew}-\r_{\nusernew}}
\end{align}
for $\Cuc{1}=\cuc{1}$, $\Cuc{2}=\cuc{2}$, \dots, $\Cuc{\he}=\cuc{\he}$, and for all non-negative $\c_{\he+1, u},\r_{\he+1, u}$ such that $\c_{\he+1, u}+\r_{\he+1, u} \leq \nslotc{\he+1}$, with $\nslotc{\he+1}= \nslotT - \he \, \tp$. Otherwise, we have
\[
\Pr  \{ \S{u}' =   \cuc{1}, \cuc{2}, \cdots, \cuc{\he+1}, \ru \} = 0 .
\] 
Hence, we initialize the $\Cuc{1}$, $\Cuc{2}$ \dots, $\Cuc{\he}$ according to the decoder state at the \ac{AP} (which is deterministically known), and $\Cuc{\he+1}$ and $\Ru$ are initialized according to a multinomial distribution, similarly as in \eqref{eq:init_cond_finite}.

Thus, $\paccessc{\he+1}$  is obtained formally as  
\[
\paccessc{\he+1} = \underset{\paccessc{\he+1}}{\arg \min} \, \per \left(\nuser, \nslotT,  u, \Omega_{\he+1,\j}',\S{u}'\right)
\] 
where we have stressed that $\per$ is a function of $\nuser$ , $\nslotT$, $u$, $\Omega_{l,\j}'$ and $\S{u}'$. Note that $\Omega_{l,\j}'$ depends implicitly on $\paccessc{\he+1}$ through \eqref{eq:de_dist_virtual}, and that $\per$ is obtained by using Theorem~\ref{theorem:class}.
Hence, it is in principle possible to run a computer search to find the value of $\paccessc{\he+1}$ which minimizes the packet error rate.
However, running such a search is computationally complex because it would require evaluating Theorem~\ref{theorem:class} multiple times\footnote{If we let $\nslot/\nuser$ be a constant, the number of unresolved users $u$ be a constant fraction of $\nuser$ and the number of slot classes $\cs$ be a constant, we have that the number of possible decoder states when $u$ users are unresolved is $\mathcal{O}(\nuser^{\cs+1})$. Furthermore, from a given decoder state when $u$ users are unresolved, the number to states that the decoder can transit to is also $\mathcal{O}(\nuser^{\cs+1})$. This is a clear indicator that the complexity of the analysis does not scale well with the number of users.}. Instead, we propose minimizing the  approximation of the packet error rate $\hat \per$ derived in Section~\ref{sec:diff_eqs_per}. This approach is suboptimal given the fact that we are minimizing an approximation of $\per$, but it makes the search for $\paccessc{\he+1}$ much faster. Thus, we obtain $\paccessc{\he+1}$  as
\[
\paccessc{\he+1} = \underset{\paccessc{\he+1}}{\arg \min} \, \hat \per \left(\nuser, \nslotT,  u, \Omega_{\he+1,\j}',\S{u}'\right)
\] 
where we remark again that that $\Omega_{l,\j}'$ depends implicitly on $\paccessc{\he+1}$ through \eqref{eq:de_dist_virtual}.

\begin{figure}[t]
        \centering
        \includegraphics[width=0.5\columnwidth]{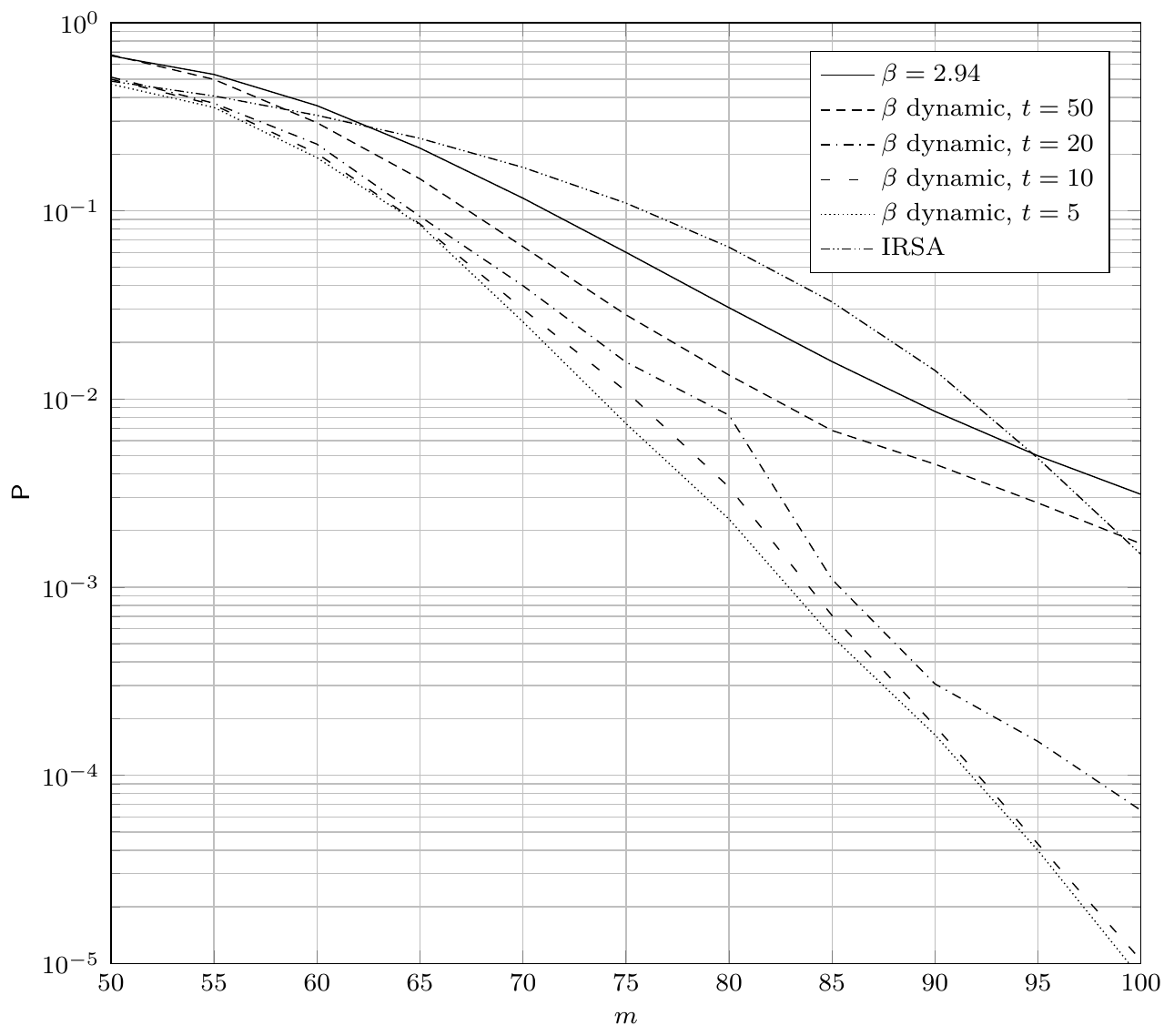}
        \caption{Packet error rate $\per$ as a function of $\nslot$ for $\nuser= 50$ users frameless ALOHA in a static and dynamic setting as well as for IRSA. Static frameless ALOHA uses $\beta=2.94$. The different curves for dynamic frameless ALOHA  represent settings in which $\beta$ is changed dynamically every $5$, $10$, $20$ and $50$ slots such that the $\hat \per$ at the end of contention-period is minimized, and the target contention period duration is set to $\nslotT=100$ slots. For IRSA the degree distribution used is $\Delta(x)=0.25x^2+0.6x^3+0.15x^8$, taken from \cite{L2011}.}
        \label{fig:per_dynamic}
\end{figure}

Fig.~\ref{fig:per_dynamic} shows the packet error rate $\per$ as a function of the duration of the contention period in slots $\nslot$, for a system with $\nuser=50$ users. Five curves are shown. The first one shows the performance of a static setting with $\beta=2.94$, which is the value that minimizes $\per$ for $\nslot=100$. 
This first curve was obtained from \eqref{eq:per} after applying the finite length analysis in  Theorem~\ref{theorem:class} for the particular case with only one slot type. 
The figure also shows the performance of a dynamic frameless ALOHA scheme in which $\beta$ is changed dynamically every $\tp$ slots in order to minimize $\hat \per$ at $\nslotT=100$. In particular, four different values of $\tp$ are considered, $5$, $10$, $20$ and $50$ slots.
The corresponding curves were obtained by means of simulations: for every value of $\nslot$, the simulation run until $250$ unsuccessful contention periods had been collected (a contention period is successful only if all $\nuser$ users are correctly decoded). Finally, the figure shows also the performance of an IRSA scheme with degree distribution  $\Delta(x)=0.25x^2+0.6x^3+0.15x^8$, taken from \cite{L2011}.
We can observe how the introduction of dynamic feedback decreases considerably the packet error rate. Furthermore, as one would expect, the performance of the dynamic scheme improves as $\tp$ decreases, that is, as the number of feedback messages increases. The curve for $\tp=20$ is particularly interesting, since one can observe how $\per$ improves after every feedback and then it degrades as time elapses; the effect is clearly visible around $\nslot=80$. In the simulated setting, it seems that sending feedback every $\tp=10$ slots may represent a good trade-off in terms of performance vs number of feedback messages sent. 
Finally, if we compare the performance of IRSA to that of frameless ALOHA, we can observe that, for the latency target of $\nslotT=100$, IRSA outperforms static frameless ALOHA, as well as dynamic frameless ALOHA with $\tp=50$, which only uses one feedback message. 
However, if we allow for more feedback messages, dynamic frameless ALOHA outperforms IRSA. For example, dynamic frameless ALOHA with $\tp=10$ outperforms IRSA by more than two orders of magnitude.

The results in Fig.~\ref{fig:per_dynamic} indicate how the introduction of feedback can greatly improve the performance of frameless ALOHA. Additional improvements of the performance could be achieved by assuming a more capable receiver, such as one that can distinguish collisions of size 2 \cite{Ogata:2019} or a receiver which can decode collisions of size $\ell>1$ \cite{lazaro2016finite}.


\section{Conclusions}\label{sec:conclusions}

In this paper we have presented a finite length analysis of multi-slot type frameless ALOHA.
The analysis is exact, but its evaluation is computationally expensive and only feasible for moderate number of users.
The analysis has been extended to derive continuous approximations of the expected ripple size and its standard deviation.
The approximations have been proven to be tight and have also been compared with simulation results.
It has also been shown how these approximations can be used to accurately estimate the packet error rate in frameless ALOHA, making it possible to analyze the performance of frameless ALOHA for large contention periods.
Finally, it has been shown how the performance of frameless ALOHA can be substantially improved by introducing feedback and adapting the slot access probability dynamically using the approximate analysis.

\begin{appendices}


\section{Proof of Theorem~\ref{theorem:class}}
\label{app:proof_theorem_class}
\begin{figure}[t]
        \centering
        \includegraphics[width=0.5\columnwidth]{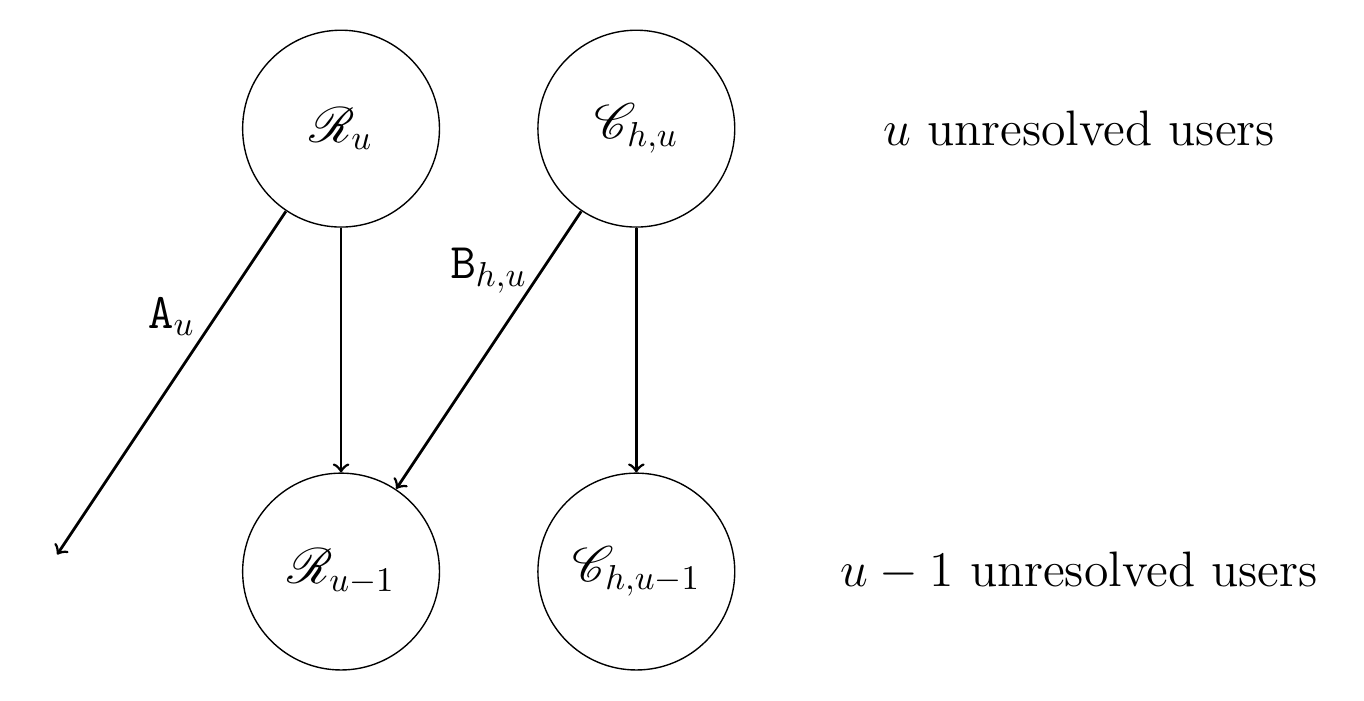}
        \caption{Evolution of the ripple and the $h$-th cloud through decoding and removal of replicas. Resolution of a single user, i.e., decoding a replica of its packet and removal of the other replicas, causes $\Erv_u$ slots to leave the ripple and $\Buc{h}$ slots to leave the $h$-th cloud and enter the ripple.} 
        \label{fig:proof}
\end{figure}

Let us start by remarking that, given the fact that every users decides whether to transmit or not in a slot independently from other users, all users are independent and statistically identical. Furthermore, all slots are mutually independent. Thus, if we look at the decoder when $u$ users are unresolved, the set of unresolved users is obtained by selecting $u$ users at random among the total of $\nuser$ users.
In particular, the proof analyzes the variation of the clouds and ripple sizes in the transition from $u$ to $u-1$ unresolved users. Since we assume  $\ru>0$, in the transition from $u$ to $u-1$ unresolved users, exactly 1 user is resolved. All the edges coming out from the resolved user are erased from the decoding graph. As a consequence, some slots might leave the cloud and enter the ripple if their reduced degree becomes one, and other slots will leave the ripple if their reduced degree decreases from 1 to 0.

Let us first focus of the number of slots leaving $\cloudc{h}{u}$ and entering $\ripple{u-1}$ in the transition, denoted by $\buc{h}$ and with associated random variable given by $\Buc{h}$. Due to the nature of frameless ALOHA, in the decoding graph the neighbor users of a slot are selected uniformly at random and without replacement\footnote{In fact, it is users who choose their neighbor slots uniformly at random and without replacement.}. Thus, random variable $\Buc{h}$ is binomially distributed with parameters $\c_{h,u}$ and $\puc{h}$, being $\puc{h}$ the probability of a generic slot $\slot$ of type $h$ leaving $\cloudc{h}{u}$ to enter $\ripple{u-1}$,
\begin{equation}
\puc{h} := \Pr \{ \slot \in \ripple{u-1} | \slot \in \cloudc{h}{u} \}= \frac { \Pr \{ \slot \in \ripple{u-1}\, , \, \slot \in \cloudc{h}{u} \} }  { \Pr \{ \slot \in \cloudc{h}{u} \}}.
\label{eq:pu_prob}
\end{equation}
We shall first focus on the numerator of \eqref{eq:pu_prob} and we shall condition it to the slot having degree $d$,
${\Pr \{ \slot \in \ripple{u-1}\, , \, \slot \in \cloudc{h}{u} | \deg(\slot)= d \}}$. This corresponds to the probability that one of the $d$ edges of  slot $\slot$ is connected to the user being resolved at the transition, one edge is connected to one of the $u-1$ unresolved users after the transition and the remaining $d-2$ edges are connected to the $\nuser-u$ resolved users before the transition. In other words, slot $\slot$ must have \emph{reduced} degree $2$ \emph{before} the transition and \emph{reduced} degree $1$ \emph{after} the transition.
It is easy to see how this probability, conditioned to $ \deg(\slot)= d$, corresponds to
\begin{align}\label{eq:z_and_l_d}
 \Pr \{ \slot \in \ripple{u-1}\, , \, & \slot \in \cloudc{h}{u} | \deg(\slot)= d \} =  
\frac{d}{\nuser} (d-1)\frac{u-1}{\nuser-1}  \frac{\binom{\nuser-u}{d-2}}{\binom{\nuser-2}{d-2}}
\end{align}
for $d\geq 2$. In the complementary case, $d < 2$, it is obvious that the slot cannot enter the ripple. Thus, for $d<2$ we have
\[
\Pr \{ \slot \in \ripple{u-1}\, , \, \slot \in \cloudc{h}{u} | \deg(\slot)= d \} = 0.
\]

Let us now concentrate on the denominator of \eqref{eq:pu_prob}, that corresponds to the probability that a slot $\slot$ is in the $h$-th cloud when $u$ users are still unresolved. This is equivalent to the probability of slot $\slot$ not being in the ripple or having reduced degree zero (all edges connected to resolved users). Hence, we have
\begin{align}
\Pr  \{ \slot \in \cloudc{h}{u}\}&=   1 -  \mathlarger{\sum}_{d=1}^{\nuser}  \Omega_{h,d}   u\frac{\binom{\nuser-u}{d-1}}{\binom{\nuser}{d}} - \mathlarger{\sum}_{d=0}^{\nuser}  \Omega_{h,d} \frac{\binom{\nuser-u}{d}}{\binom{\nuser}{d}}
\label{eq:z}
\end{align}
where the first summation on the right hand side corresponds to the probability of a slot being the ripple, and the second summation corresponds to the probability of a slot having reduced degree zero.
Inserting \eqref{eq:z_and_l_d} and \eqref{eq:z} in \eqref{eq:pu_prob}, the expression of $\puc{h}$ in \eqref{eq:pu_theorem_class} is obtained, and the variation in size of the clouds is determined, i.e., the \acp{PMF} of the random variables $\Buc{h}$ are obtained.

We focus next on the variation in size of the ripple in the transition from $u$ to $u-1$ unresolved users.
In this transition, some slots enter the ripple (in total $\sum_{h}\buc{h}$ slots), but there are also slots leaving the ripple.
Let us denote by $\erv_u$ the number of slots leaving the ripple in the transition from $u$ to $u-1$ unresolved users, and let us refer to the associated random variable as $\Erv_u$. Assuming that the ripple is not empty\footnote{If the ripple is empty, $\ru=0$, no slots can  leave the ripple. Moreover, decoding stops, so there is no transition.}, the decoder will select uniformly at random one slot from the ripple, that we denote as $\slot$. The only neighbour of $\slot$, $c$ will get resolved and all slots in the ripple that are connected to $c$ will leave the ripple in the transition.  Hence, slot $\slot$ will leave the ripple, and, additionally, the remaining $\ru-1$ slots in the ripple will leave the ripple  independently with  probability $1/u$, which is the probability that they have $c$ as neighbour. Thus, the probability mass function of $\Erv_u$ is given by
\begin{align}
\Pr\{\Erv_u=\erv_u & |\Ru=\ru\}=
\binom{\ru-1}{\erv_u-1} \left(\frac{1}{u}\right)^{\erv_u-1} \mkern-3mu \left( 1- \frac{1}{u} \right)^{\ru-\erv_u}.
\end{align}

Note that, by definition, we have that
${\r_{u-1}= \ru -\erv_u +  \sum_{h=1}^{\cs} \buc{h}} $ 
and 
${\c_{h,u-1} = \cuc{h} - \buc{h}}$.
Hence, the probability of transiting from state
$\s{u} =(\cuc{1}, \cuc{2}, \cdots, \cuc{\cs}, \ru ) $
to state $\s{u} + \w$, with $\w$ given by
\[{\w =(-\buc{1}, -\buc{2}, \cdots , -\buc{\cs}, \sum_{h=1}^{\cs} \buc{h} - \erv_u ) }\] corresponds to 
\begin{align}\label{eq:joint}
\Pr \{\Buc{1}=\buc{1}, \Buc{2}=\buc{2}, \cdots, \Buc{\cs}=\buc{\cs}, \Erv_u = \erv_u\}.
\end{align}
The proof is finalized by observing that the joint distribution in \eqref{eq:joint}, due to independence of $\Buc{h}$, $h=1,2,\cdots, \cs$, and $\Erv_u$, is obtained as the product of the \acp{PMF} of the individual random variables, which yields \eqref{eq:prob_transition}.

\vspace{-0.8cm}
\hspace{15cm}$\blacksquare$

\section {Proof of Theorem~\ref{theorem_state_gen_fun}} \label{sec:theorem_state_proof}

We provide a proof only for the case of $\cs=2$  slot types. The proof follows closely \cite[Theorem~1]{shokrollahi2009theoryraptor}.  By definition, the state generating function of the frameless ALOHA decoder when $u-1$ users are undecoded can be written as
\begin{equation} \label{eq:def_state_gen}
\P_{u-1} (\xc{1},\xc{2},\y) = \sum_{ \substack{ \cc{1}'\geq 0 \\ \cc{2}' \geq 0 \\  r' \geq 1 }} \pccruprime \xc{1}^{\cc{1}'} \, \xc{2}^{\cc{2}'} \,\y^{r'-1}
\end{equation}
Our objective is expressing $\P_{u-1} (\xc{1},\xc{2},\y) $ as a function of $\P_{u} (\xc{1},\xc{2},\y)$. For this purpose, we must consider the contribution of all previous states
$\s{u} =(\cuc{1}, \cuc{2}, \ru ) $ to the future state ${\s{u-1}=( \cucs{1}{u-1}',\cucs{2}{u-1}', \r_{u-1}')}$. In the remainder of the proof, we shall drop the subscript $u$ for the sake of readability.
As done in the proof of Theorem~\ref{theorem:class}, we denote by $\bh{h}$ the number of slots leaving $\Cuc{h}$ in the transition from $u$ to $u-1$ undecoded slots. Furthermore, we denote by $\erv$ the number of slots leaving the ripple in the transition. Obviously, we have
\begin{align}
\cc{1} &= \cc{1}' + \bh{1} \\
\cc{2} &= \cc{2}' + \bh{2} \\
\r &= \r' + \erv - \bh{1} - \bh{2}.
\end{align}
Thus, we can rewrite \eqref{eq:def_state_gen} as follows
\begin{align}
\P_{u-1} &(\xc{1},\xc{2},\y) = \frac{1}{\y} \sum_{ \substack{ \cc{1} \geq 0 \\ \cc{2}  \geq 0 \\  r  \geq 1 }}   \pccru
\sum_{ \substack{ 1 \leq \erv \leq r \\ 0 \leq \bh{1} \leq \cc{1} \\  0 \leq \bh{2} \leq \cc{2}  \\  r-a + \bh{1} + \bh{2} > 0 }}  
 \xc{1}^{\cc{1}-\bh{1}} \, \xc{2}^{\cc{2}-\bh{2}} \,\y^{r-\erv + \bh{1} + \bh{2}}  \\
& \times  \binom{\cc{1}}{\bh{1}} \puc{1}^{ \bh{1} }   (1-\puc{1})^{ \cc{1} - \bh{1} }
\binom{\cc{2}}{\bh{2}} \puc{2}^{ \bh{2} }   (1-\puc{2})^{ \cc{2} - \bh{2} } 
\binom{\r-1}{\erv-1} \left(\frac{1}{u}\right)^{\erv -1} \left(1- \frac{1}{u}\right)^{\r - \erv }.
\end{align}
By grouping different terms together, we obtain
\begin{align}
\P_{u-1} (\xc{1},\xc{2},\y) &= \frac{1}{\y} \sum_{ \substack{ \cc{1} \geq 0 \\ \cc{2}  \geq 0 \\  r  \geq 1 }}  \pccru
\sum_{ \substack{ 1 \leq \erv \leq r \\ 0 \leq \bh{1} \leq \cc{1} \\  0 \leq \bh{2} \leq \cc{2}  \\  r-a + \bh{1} + \bh{2} > 0 }}  
\binom{\cc{1}}{\bh{1}} ( \y \, \puc{1} )^{ \bh{1} }   \left( \xc{1} (1-\puc{1}) \right)^{ \cc{1} - \bh{1} } \\
& \times  \binom{\cc{2}}{\bh{2}} ( \y \, \puc{2} )^{ \bh{2} }   \left( \xc{2} (1-\puc{2}) \right)^{ \cc{2} - \bh{2} } 
\binom{\r-1}{\erv-1} \left(\frac{1}{u}\right)^{\erv -1} \left( \y \left(1- \frac{1}{u}\right)\right)^{\r - \erv }.
\end{align}
We now take out the term $\erv=r$ from the summation. In case that $\erv=r$, $\bh{1}$ and $\bh{2}$ cannot be both zero at the same time. Thus, we have
\begin{align}
\label{eq:grouped}
\P_{u-1} (\xc{1},\xc{2},\y) &=  \frac{1}{\y} \sum_{ \substack{ \cc{1} \geq 0 \\ \cc{2}  \geq 0 \\  r  \geq 1 }} \pccru 
\Bigg( \sum_{ \substack{ 1 \leq \erv < r \\ 0 \leq \bh{1} \leq \cc{1} \\  0 \leq \bh{2} \leq \cc{2}  }}
 \binom{\cc{1}}{\bh{1}} ( \y \, \puc{1} )^{ \bh{1} }   \left( \xc{1} (1-\puc{1}) \right)^{ \cc{1} - \bh{1} } \\
& \times  \binom{\cc{2}}{\bh{2}} ( \y \, \puc{2} )^{ \bh{2} }   \left( \xc{2} (1-\puc{2}) \right)^{ \cc{2} - \bh{2} } 
\binom{\r-1}{\erv-1} \left(\frac{1}{u}\right)^{\erv -1} \left( \y \left(1- \frac{1}{u}\right)\right)^{\r - \erv } \Bigg) \\
& + \left( \frac{1}{u} \right)^{\r-1} \Bigg( \sum_{ \substack{ 0 \leq \bh{1} \leq \cc{1} \\  1 \leq \bh{2} \leq \cc{2}  }}
\binom{\cc{1}}{\bh{1}} ( \y \, \puc{1} )^{ \bh{1} }   \left( \xc{1} (1-\puc{1}) \right)^{ \cc{1} - \bh{1} } \\
& \times
\binom{\cc{2}}{\bh{2}} ( \y \, \puc{2} )^{ \bh{2} }   \left( \xc{2} (1-\puc{2}) \right)^{ \cc{2} - \bh{2} }\\
&+ \left( \xc{2} (1-\puc{2}) \right)^{ \cc{2}} \mkern-15mu \sum_{ \substack{ 1 \leq \bh{1} \leq \cc{1} }}  \mkern-15mu
\binom{\cc{1}}{\bh{1}} ( \y \, \puc{1} )^{ \bh{1} }   \left( \xc{1} (1-\puc{1}) \right)^{ \cc{1} - \bh{1} } \Bigg)
\end{align}

By using in \eqref{eq:grouped} the results of the following two finite sums
\begin{align}
  &\sum_{\bh{1}=0}^{\cc{1} } \binom{\cc{1}}{\bh{1}} ( \y \, \puc{1} )^{ \bh{1} }  \left( \xc{1} (1-\puc{1}) \right)^{ \cc{1} - \bh{1} } 
  = (\y \, \puc{1} +  \xc{1} (1- \puc{1}))^{\cc{1}}
\end{align}
\begin{align}
  &\sum_{\erv=1}^{\r-1}  \binom{\r-1}{\erv-1} \left(\frac{1}{u}\right)^{\erv -1} \left( \y \left(1- \frac{1}{u}\right)\right)^{\r - \erv }  
  = \left( \frac{1}{u} + \y (1- \frac{1}{u}) \right)^{\r-1} - \left( \frac{1}{u}\right)^{\r-1}
\end{align}
after some manipulation, we obtain
\begin{align}
\P_{u-1} (\xc{1},\xc{2},\y) & =   (\y \, \puc{1} +  \xc{1} (1- \puc{1}))^{\cc{1}}  
(\y \, \puc{2} +  \xc{2} (1- \puc{2}))^{\cc{2}} \\
& \times \left( \frac{1}{u} + \y (1- \frac{1}{u}) \right)^{\r-1} 
- \left( \frac{1}{u}\right)^{\r-1}  \left( \xc{1} (1-\puc{1}) \right)^{ \cc{1}} \left( \xc{2} (1-\puc{2}) \right)^{ \cc{2}}
\end{align}
which yields \eqref{eq:theorem_new}.

In order to complete the proof, we need to provide a proof for \eqref{eq:p_k}, which specifies the initial condition of the recursive expression given in \eqref{eq:theorem_new}. By definition, $\pccrn$ is simply the probability that before decoding starts ($u=\nuser$) we have exactly $\cc{1}$ slots in the first cloud, $\cc{2}$ slots in the second cloud and $r$ slots in the ripple.
Let us decompose $\r$ as $\r = \r_1 + \r_2$, , with $\r_1 \geq 0$ and $\r_2 \geq 0$, where  $\r_1$ and $\r_2$ represent the number of slots of type 1 and type 2 in the ripple, respectively. We have
\begin{align}\label{eq:pcrn_def}
\P_{\nuser} (\xc{1},\xc{2},\y)  & = \sum_{ \substack{ \validstates }}  \pccrn \xc{1}^{\cc{1}}  \xc{2}^{\cc{2}} \y^{r-1}  = \sum_{ \substack{ \validstates'}}  \pccrrn \xc{1}^{\cc{1}}  \xc{2}^{\cc{2}} \y^{\r_1+\r_2-1} 
\end{align}
with 
\begin{align} 
\validstates'= \Big\{ &(\cc{1}, \cc{2}, \r_1 , \r_2) | \cc{1} \geq 0,  \cc{2} \geq 0, \r_1 \geq 0, \r_2 \geq 0, \r_1+\r_2 \geq 1, \\
& \cc{1} + \r_1 \leq \nslotc{1}, \cc{2} + \r_2 \leq \nslotc{2} \Big\}
\end{align}
where we remark that though $\r_1$ and $\r_2$ can take value $0$, they cannot do so at the same time, since otherwise decoding would not be able to start. \newline
From \eqref{eq:init_cond_finite}, we have that
\begin{align}\label{eq:valid_states_prime}
  \pccrrn & = \prod_{h=1}^{2}\frac{\nslotc{h}!}{\c_{h}! \, \r_{h}!\, (\nslotc{h}-\c_{h}-\r_{h})!}   
  \left( 1-\Omega_{h,1}- \Omega_{h,0}\right)^{\c_{h}}\, {\Omega_{h,1}}^{\r_{\nuser}} \, {\Omega_{h,0}}^{\nslotc{h}-\c_{h}-\r_{h}}.
\end{align}
If we replace this expression into the right hand side of \eqref{eq:pcrn_def},  compute the summation over the set of all valid states $\validstates'$ given in \eqref{eq:valid_states_prime}, and make use of the multinomial Theorem, one obtains \eqref{eq:p_k}. 
This completes the proof for $\cs=2$. The proof for generic $\cs$ uses the same reasoning and follows exactly the same steps.

\vspace{-0.75cm}
\hspace{15cm}$\blacksquare$

\section{Initial Conditions for Dynamic Feedback}
Let us first consider the clouds, $C_h(\nusernew)$.  For $h \leq \he$
\begin{align}
C_h(\nusernew) &= \sum_{ \validstates }  \cc{h} \, \pcccrn =
\cuc{h}
\left( 1 - \big(1-\Omega_{\he+1,1} \big)^{\nslotc{\he+1}}\right)
\end{align}
thus, we have
\[
\cho = \frac{ \cuc{h} }{ \nslotnew}
\frac{\left( 1 - \big(1-\Omega_{\he+1,1} \big)^{\nslotc{\he+1}}\right)}{1-\Omega_{h,1} - \Omega_{h,0}}.
\]
For $h=\he+1$ we have
\[
C_h(\nusernew)=
\nslotc{h} \left( 1 - \Omega_{h,0} - \Omega_{h,1} \right)  \left( 1 - \big(1-\Omega_{h,1} \big)^{\nslotc{h}}\right)
\]
which yields
\[
\cho = \frac{\nslotc{h}}{\nslot} \left( 1 - \big(1-\Omega_{h,1} \big)^{\nslotc{h}}\right).
\]

For the ripple we have
\begin{align}
R(\nusernew) &=  \sum_{ \validstates} (\r-1) \, \pcccrn 
= \sum_{h=1}^{\he} \cuc{h} +  \nslotc{\he+1}\Omega_{\he+1,1} - 1 +  \big( 1- \Omega_{\he+1,1}\big)^{\nslotc{\he+1}}
\end{align}
hence, imposing $\hat R(x=1) = R(\nusernew)/ \nslotnew$ yields
\begin{align}\label{eq:r0_dyn}
\ro &=
\sum_{h=1}^{\he} \frac{ \cuc{h} }{\nslotnew} \left( 1 - \frac{\left( 1 - \big(1-\Omega_{\he+1,1} \big)^{\nslotc{\he+1}}\right)}{1-\Omega_{h,1} - \Omega_{h,0}} \right)+  \frac{ \nslotc{\he+1} }{\nslotnew} \Omega_{\he+1,1}
- \frac{1}{\nslotnew} \left( 1 -   \big( 1- \Omega_{\he+1,1}\big)^{\nslotc{\he+1}} \right).
\end{align}

\end{appendices}

\section*{Acknowledgments}
The authors would like to thank Gianluigi Liva for pointing out the possibility of applying the analysis tools of LT codes to frameless ALOHA.


\begin{thebibliography}{10}
\providecommand{\url}[1]{#1}
\csname url@samestyle\endcsname
\providecommand{\newblock}{\relax}
\providecommand{\bibinfo}[2]{#2}
\providecommand{\BIBentrySTDinterwordspacing}{\spaceskip=0pt\relax}
\providecommand{\BIBentryALTinterwordstretchfactor}{4}
\providecommand{\BIBentryALTinterwordspacing}{\spaceskip=\fontdimen2\font plus
\BIBentryALTinterwordstretchfactor\fontdimen3\font minus
  \fontdimen4\font\relax}
\providecommand{\BIBforeignlanguage}[2]{{%
\expandafter\ifx\csname l@#1\endcsname\relax
\typeout{** WARNING: IEEEtran.bst: No hyphenation pattern has been}%
\typeout{** loaded for the language `#1'. Using the pattern for}%
\typeout{** the default language instead.}%
\else
\language=\csname l@#1\endcsname
\fi
#2}}
\providecommand{\BIBdecl}{\relax}
\BIBdecl

\bibitem{lazaro:SCC2017}
F.~L{\'a}zaro and C.~Stefanovic, ``Finite-length analysis of frameless
  {ALOHA},'' in \emph{Proc. 11th International ITG Conf. on Sys., Commun. and
  Coding}, Hamburg, Germany, Feb. 2017.

\bibitem{stefanovic:globecom2017}
C.~Stefanovic, F.~L{\'a}zaro, and P.~Popovski, ``Frameless {ALOHA} with
  reliability-latency guarantees,'' in \emph{Proc. of IEEE Global Commun.
  Conf.}, Singapore, Dec. 2017, pp. 1--6.

\bibitem{TS36.321}
{3GPP}, ``{3GPP TS 36.321 V15.4.0}: {M}edium access control protocol
  specification; ({R}elease 14),'' Tech. Rep., Jan. 2019.

\bibitem{R1-1808304}
3GPP, ``{D}iscussion on the reliability enhancement for grant-free
  transmission,'' Tech. Rep., Aug. 2018.

\bibitem{Laya2014}
A.~Laya, L.~Alonso, and J.~Alonso-Zarate, ``Is the random access channel of
  {LTE} and {LTE-A} suitable for {M2M} communications? {A} survey of
  alternatives,'' \emph{IEEE Commun. Surv. Tut.}, vol.~16, no.~1, pp. 4--16,
  First Quarter 2014.

\bibitem{R1975}
L.~G. Roberts, ``{ALOHA} packet system with and without slots and capture,''
  \emph{ACM SIGCOMM Comput. Commun. Rev.}, vol.~5, no.~2, pp. 28--42, Apr.
  1975.

\bibitem{CGH2007}
E.~Cassini, R.~D. Gaudenzi, and O.~del Rio~Herrero, ``{C}ontention resolution
  diversity slotted {ALOHA} {(CRDSA)}: {A}n enhanced random access scheme for
  satellite access packet networks,'' \emph{{IEEE} Trans. Wireless Commun.},
  vol.~6, no.~4, pp. 1408--1419, Apr. 2007.

\bibitem{L2011}
G.~Liva, ``{G}raph-based analysis and optimization of contention resolution
  diversity slotted {ALOHA},'' \emph{{IEEE} Trans. Commun.}, vol.~59, no.~2,
  pp. 477--487, Feb. 2011.

\bibitem{PLC2011}
E.~Paolini, G.~Liva, and M.~Chiani, ``{H}igh throughput random access via codes
  on graphs: coded slotted {ALOHA},'' in \emph{Proc. of IEEE Int. Conf.
  Commun.}, Kyoto, Japan, Jun. 2011.

\bibitem{LPLC2012}
G.~Liva, E.~Paolini, M.~Lentmaier, and M.~Chiani, ``{S}patially-coupled random
  access on graphs,'' in \emph{Proc. of IEEE Int. Symp. Inf. Theory}, Boston,
  MA, USA, Jul. 2012.

\bibitem{SPV2012}
C.~Stefanovic, P.~Popovski, and D.~Vukobratovic, ``{F}rameless {ALOHA} protocol
  for wireless networks,'' \emph{{IEEE} Commun. Lett.}, vol.~16, no.~12, pp.
  2087--2090, Dec. 2012.

\bibitem{PSLP2014}
E.~Paolini, C.~Stefanovic, G.~Liva, and P.~Popovski, ``Coded random access:
  applying codes on graphs to design random access protocols,'' \emph{IEEE
  Commun. Mag.}, vol.~53, no.~6, pp. 144--150, Jun. 2015.

\bibitem{JBVC2015}
D.~{Jakoveti\' c}, D.~{Bajovi\' c}, D.~{Vukobratovi\' c}, and V.~{Crnojevi\'
  c}, ``Cooperative slotted {ALOHA} for multi-base station systems,''
  \emph{{IEEE} Trans. Commun.}, vol.~63, no.~4, pp. 1443--1456, Apr. 2015.

\bibitem{SGB2017}
E.~{Sandgren}, A.~{Graell i Amat}, and F.~{Br\" annstr\" om}, ``On frame
  asynchronous coded slotted {ALOHA}: asymptotic, finite length, and delay
  analysis,'' \emph{{IEEE} Trans. Commun.}, vol.~65, no.~2, pp. 691--704, Feb.
  2017.

\bibitem{SP2013}
C.~Stefanovic and P.~Popovski, ``{ALOHA} random access that operates as a
  rateless code,'' \emph{{IEEE} Trans. Commun.}, vol.~61, no.~11, pp.
  4653--4662, Nov. 2013.

\bibitem{luby02:LT}
M.~Luby, ``{LT} codes,'' in \emph{Proc. 43rd Annu. IEEE Symp. on Found. of
  Comp. Science}, Vancouver, Canada, Nov. 2002, pp. 271--282.

\bibitem{OIA2018}
S.~{Ogata}, K.~{Ishibashi}, and G.~T.~F. {de Abreu}, ``Optimized frameless
  {ALOHA} for cooperative base stations with overlapped coverage areas,''
  \emph{{IEEE} Trans. Wireless Commun.}, vol.~17, no.~11, pp. 7486--7499, Nov.
  2018.

\bibitem{Ogata:2019}
S.~{Ogata} and K.~{Ishibashi}, ``Application of zigzag decoding in frameless
  {ALOHA},'' \emph{IEEE Access}, vol.~7, pp. 39\,528--39\,538, Apr. 2019.

\bibitem{ivanov:floor}
M.~Ivanov, F.~Br\"annstr\"om, A.~{Graell i Amat}, and P.~Popovski, ``Error
  floor analysis of coded slotted {ALOHA} over packet erasure channels,''
  \emph{{IEEE} Commun. Lett.}, vol.~19, no.~3, pp. 419--422, Mar. 2015.

\bibitem{graell:2018}
A.~G. i~Amat and G.~Liva, ``Finite-length analysis of irregular repetition
  slotted {ALOHA} in the waterfall region,'' \emph{{IEEE} Commun. Lett.},
  vol.~22, no.~5, pp. 886--889, May 2018.

\bibitem{fereydouniannon}
M.~Fereydounian, X.~Chen, H.~Hassani, and S.~S. Bidokhti, ``Non-asymptotic
  coded slotted {ALOHA},'' in \emph{Proc. of IEEE Int. Symp. Inf. Theory},
  Paris, France, Jul. 2019, pp. 111--115.

\bibitem{MGS2018}
A.~{Mengali}, R.~{De Gaudenzi}, and {\v C}.~{Stefanovi\' c}, ``On the modeling
  and performance assessment of random access with {SIC},'' \emph{{IEEE} J.
  Select. Areas Commun.}, vol.~36, no.~2, pp. 292--303, Feb. 2018.

\bibitem{TR38.913}
{3GPP}, ``{3GPP TR 38.913 V15.0.0}: {S}tudy on scenarios and requirements for
  next generation access technologies; ({R}elease 15),'' Tech. Rep., Jun. 2018.

\bibitem{Karp2004}
R.~Karp, M.~Luby, and A.~Shokrollahi, ``Finite length analysis of {LT} codes,''
  in \emph{Proc. IEEE Int. Symp. Inf. Theory}, {Chicago, IL, USA}, Jun. 2004.

\bibitem{lazaro:inact2017}
F.~L{\'a}zaro, G.~Liva, and G.~Bauch, ``Inactivation decoding of {LT} and
  {Raptor} codes: {A}nalysis and code design,'' \emph{{IEEE} Trans. Commun.},
  vol.~65, no.~10, pp. 4114--4127, Oct. 2017.

\bibitem{shokrollahi2009theoryraptor}
A.~Shokrollahi, ``Theory and applications of {Raptor} codes,'' \emph{Mathknow},
  vol.~3, pp. 59--89, 2009.

\bibitem{Maatouk:2012}
G.~Maatouk and A.~Shokrollahi, ``Analysis of the second moment of the {LT}
  decoder,'' \emph{{IEEE} Trans. Inf. Theory}, vol.~58, no.~5, May 2012.

\bibitem{Maatouk:phd}
G.~Maatouk, ``Graph-based codes and generalized product constructions,'' Ph.D.
  dissertation, \'Ecole Polytechnique F\'ed\'erale de Lausanne, Switzerland,
  August 2013.

\bibitem{lazaro2016finite}
F.~L\'azaro and {\v{C}}.~Stefanovi{\'c}, ``Finite-length analysis of frameless
  {ALOHA} with multi-user detection,'' \emph{{IEEE} Commun. Lett.}, vol.~21,
  no.~4, pp. 769--772, 2016.

\end{thebibliography}

\end{document}